\documentclass[preprint2]{aastex}
\usepackage{lscape}
\usepackage{breqn}

\shorttitle{High energy pulsations from PSR J1813-1246}
\shortauthors{Marelli et al.}

\begin{document}

\title{On the puzzling high-energy pulsations of the energetic radio-quiet $\gamma$-ray pulsar J1813$-$1246}

\author{M. Marelli\textsuperscript{1}, A. Harding\textsuperscript{2}, D. Pizzocaro\textsuperscript{1,3}, A. De Luca\textsuperscript{1},
K.S. Wood\textsuperscript{4},
P. Caraveo\textsuperscript{1}, D. Salvetti\textsuperscript{1}, P. M. Saz Parkinson\textsuperscript{5,6}, F. Acero\textsuperscript{7}}

\footnote{INAF - Istituto di Astrofisica Spaziale e Fisica Cosmica Milano, via E. Bassini 15, 20133 Milano, Italy\\
\textsuperscript{2}Astrophysics Science Division, NASA/Goddard Space Flight Center, Greenbelt, MD 20771, USA\\
\textsuperscript{3}Universit\'a degli Studi dell'Insubria - Via Ravasi 2 - 21100 Varese - Italy\\
\textsuperscript{4}Space Science Division, Naval Research Laboratory, Washington DC 20375\\
\textsuperscript{5}Department of Physics, The University of Hong Kong, Pokfulam Road, Hong Kong\\
\textsuperscript{6}Santa Cruz Institute for Particle Physics, University of California, Santa Cruz, CA 95064\\
\textsuperscript{7}Laboratoire AIM, CEA-IRFU/CNRS/Université Paris Diderot, Service d’Astrophysique,
CEA Saclay, 91191 Gif sur Yvette, France}
\email{marelli@lambrate.inaf.it}

{\bf Abstract}

We have analyzed the new deep {\it XMM-Newton} and {\it Chandra} observations of the energetic radio-quiet pulsar J1813$-$1246.
The X-ray spectrum is non-thermal, very hard and absorbed. Based on spectral considerations,
we propose that J1813 is located at a distance further than 2.5 kpc.
J1813 is highly pulsed in the X-ray domain, with a light curve characterized by two sharp, asymmetrical peaks,
separated by 0.5 in phase. We detected no significant X-ray spectral changes during the pulsar phase.
We extended the available {\it Fermi} ephemeris to five years. We found two glitches. The $\gamma$-ray lightcurve
is characterized by two peaks, separated by 0.5 in phase, with a bridge in between and no off-pulse emission.
The spectrum shows clear evolution in phase, being softer at the peaks and hardenning towards the bridge.
The X-ray peaks lag the $\gamma$-ray ones by 0.25 in phase.
We found a hint of detection in the 30-500 keV band with {\it INTEGRAL} IBIS/ISGRI, that is consistent with the 
extrapolation of both the soft X-ray and $\gamma$-ray emission of J1813.
The peculiar X and $\gamma$-ray phasing suggests a singular emission geometry. We discuss some
possibilities within the current pulsar emission models. Finally, we develop an alternative geometrical model
where the X-ray emission comes from polar cap pair cascades.

\section{Introduction}

The Large Area Telescope (LAT) on board the {\it Fermi Gamma-ray Space
Telescope} (hereafter, $Fermi$-LAT) is providing new insights into the $\gamma$-ray
pulsar population, revolutionizing our understanding of pulsar high-energy emission \citep{car13}.
The wealth of detections \citep{abd13} confirms the importance of the $\gamma$-ray channel
in the overall energy budget of rotation-powered pulsars and paves the way for a better
understanding of the three-dimensional structure and electrodynamics of neutron star magnetospheres.
Indeed, radio and $\gamma$-ray light curves are shaped by the geometry as well as by the
emission processes at work in pulsar magnetospheres \citep[see e.g.][]{wat11,pie12,pie14}.
Based on the phenomenology of $\sim$150 $\gamma$-ray detections, models with emission originating at high
altitudes in the magnetosphere \citep[e.g. outer and slot-gap,][]{che86,har04} are favored over models with near-surface emission
\citep[e.g. polar cap,][]{har13}.

Fitting $\gamma$-ray and radio light curves simultaneously is a
promising way to constrain geometric parameters of the pulsar \citep[e.g.,][]{pie14}. 
Exploiting the (magnetospheric) non-thermal pulsar X-ray light curves could further improve such an approach,
adding another piece to the pulsar emission puzzle. This would also allow the localization of 
the emitting region(s) responsible for the non-thermal pulsed X-ray emission with respect to the high altitude $\gamma$-ray emitting one(s).\\
Pulsar X-ray light curves are very diverse, with one or more peaks, broad or narrow, and
a range of phase lags between radio, $\gamma$-ray, and X-ray peaks. 
Indeed, with the notable exception of the Crab pulsar (among the young ones), the multi-wavelength behavior of isolated neutron stars is 
complex, with radio, optical, X and $\gamma$-ray light curves 
usually misaligned, pointing to different, and currently unknown, emitting regions in the pulsar magnetosphere.
The rich X-ray phenomenology has not yet been fully exploited, leaving a number of open questions. 

Here we report the results of deep joint {\it XMM-Newton} and {\it Chandra} observations aimed at
searching for pulsations and performing a phase-resolved spectral analysis of
the radio-quiet {\it Fermi}-LAT pulsar J1813$-$1246 (hereafter, J1813)
in the soft X-ray band (0.3-10 keV). Our 
X-ray observations also enable us to study the possible extended emission from its pulsar wind nebula (PWN).\\

J1813 was discovered within a few months of the launch of {\it Fermi},
in a blind pulsation search of LAT data \citep{abd09}. It is one of
the brightest $\gamma$-ray pulsars, making it into the {\it Fermi}-LAT
Bright Source List as 0FGL J1813.5-1248 \citep{abd09b}.
Its period P $\sim$ 48.1 ms and period derivative $\dot{P}$ $\sim1.76\times10^{-14}$ s s$^{-1}$
point to a spin-down energy loss rate $\dot{E}$ $\sim$ 6.24$\times10^{36}$ erg s$^{-1}$
and characteristic age $\tau_c$ = 43 kyr, making it
the fastest-spinning known radio-quiet pulsar and the second most energetic one \citep[see ][]{abd13}.
Its $\gamma$-ray light curve exhibits two fairly broad peaks 180$^{\circ}$ apart (peak phase separation of 0.49$\pm$0.01),
with a clear asymmetric bridge emission \citep{abd13}.
Although there is no reliable distance measurement for J1813, its pseudo-distance,
which hinges on the observed correlation between intrinsic $\gamma$-ray luminosity and
$\dot{E}$ \citep{saz10}, is $\sim$1.5 kpc (this would result in a $\gamma$-ray
efficiency of $\sim$0.01, typical of energetic $\gamma$-ray pulsars).
This pulsar exhibited a glitch around 2009 September 20 \citep{ray11}.
Despite dedicated Green Bank Telescope radio observations at 0.82 and 2 GHz
no radio emission was detected down to a 17$\mu$Jy limit \citep{abd13}.
A possible X-ray counterpart was detected by {\it Swift} soon after
the discovery of the pulsar \citep{abd09}, and was confirmed to be
coincident with the precise {\it Fermi}-LAT timing position
\citep{ray11}. The bright ($10^{-12}$ erg cm$^{-2}$ s$^{-1}$) X-ray counterpart
unveiled by {\it Swift} \citep{mar11} was later confirmed by {\it
  Suzaku} \citep{abd13}, possibly associated with a nebular emission extending up to a few tens of arcseconds.

\section{Observations and data reduction}

Our deep {\it XMM-Newton} observation of J1813 was performed on 2013 March 10
and lasted 108.9 ks. The PN camera \citep{str01} of the EPIC instrument was operating
in Small Window mode, with a time resolution of 5.6 ms over a 4' $\times$ 4' Field Of View (FOV),
while the Metal Oxide Semi-conductor (MOS) detectors \citep{tur01} were set in Full Frame mode (2.6 s time resolution on
a 15' radius FOV). The thin optical filter was used for the PN and the medium filter for the MOS cameras.
We used the {\it XMM-Newton} Science
Analysis Software (SAS) v13.0. We performed a standard analysis of
high particle background \citep[following ][]{del05}. We cross-checked
the results with the SAS tool {\tt bkgoptrate} (also used for the 3XMM
source
catalog\footnote{\url{http://xmmssc-www.star.le.ac.uk/Catalogue/xcat\_public\_3XMM-DR4.html}}). This
tool searches for the point at which the maximum signal-to-noise (S/N)
ratio is achieved for the given background time series after the bins above a threshold are excluded.
Both analyses revealed no significant contamination from flares. We
selected 0-4 pattern events for PN and 0-12 for the MOS detectors in
the 0.3-10 keV energy range, following \citet{mar13}. Then, we excluded the 0.3-0.4 keV energy range
for the PN owing to the presence of bright columns.
Due to the high degree of absorption in our source, the number of expected counts in the 0.3-0.4 keV energy range is negligible
($10^{-12}$ counts s$^{-1}$ for the best fit spectrum, obtained using the WebPimms HEASARC tool).
For each spectrum we generated ad hoc response matrices and effective area files using the SAS tools {\tt rmfgen} and {\tt arfgen}.\\
To fully characterize both the pulsar and its putative nebula, we also obtained a {\it Chandra/ACIS-S} \citep{gar03} exposure
of the field. The observation was performed on 2013 July 22 and lasted 50.4 ks.
The telemetry mode was set to Very Faint, recommended in order to reduce background in extended sources.
We used the {\it Chandra} Interactive Analysis of Observation (CIAO) software v4.5.
We also re-analyzed a public 25.7ks {\it Suzaku} \citep{mit07} observation performed on 2010 March 22.
The HEAsoft package (v6.15) was used to analyze {\it Suzaku}
data, following the standard recommendations \footnote{\url{http://heasarc.nasa.gov/docs/suzaku/aehp\_data\_analysis.html}}.

\section{$\gamma$-ray analysis} \label{gray}

The {\it Fermi}-LAT dataset we used to extend the $\gamma$-ray ephemeris of J1813 spans five years, from 2008
August 4 to 2013 August 4. P7REP {\tt Source} class events were selected with reconstructed energies from 0.1 to 100 GeV
and with arrival directions within 20$^{\circ}$ of the source position.
We excluded $\gamma$-rays collected when the LAT was not in nominal science operations mode, 
when the spacecraft rocking angle exceeded 52$^{\circ}$, or when the Sun was within 5$^{\circ}$
of the pulsar position. Moreover, to reduce contamination by residual
$\gamma$ rays from the bright limb of the Earth, we excluded photons with measured zenith angles $>100^{\circ}$.
We performed a binned maximum likelihood analysis, following \citet{abd13}.
We used the {\it Fermi} Science tools v09r32p04, Instrument Response Functions P7REP\_SOURCE\_V15,
the Galactic and isotropic models obtained by the LAT collaboration from the analysis of four years of
data\footnote{\url{http://fermi.gsfc.nasa.gov/ssc/data/access/lat/BackgroundModels.html}}.
The analysis tools, instrument response functions, and diffuse emission models are available from the {\it Fermi} Science Support Center\footnote{\url{http://fermi.gsfc.nasa.gov/ssc}}.
The source models were taken from the two-year source and pulsar catalogs \citep{nol12,abd13}.
In our model of the region, post-fit spatial residuals did not  reveal the need
for any additional source, beyond those in the two-year catalog.
The pulsar $\gamma$-ray spectrum is consistent with a power law with
an exponential cutoff with $\Gamma$ = 2.15$\pm$0.02 and cutoff energy E$_c$ = 3.6$\pm$0.3 GeV (1$\sigma$ errors).
These results are in agreement with those in \citet{abd13}.

Since \citet{ker11} reports an increase in the sensitivity
to pulsations by more than 50\%, under a wide range of conditions,
when using photon weighting techniques on {\it Fermi}-LAT sources,
we used the {\it Fermi} Science Tool {\tt gtsrcprob}, that combines the spectral results with
the energy-dependent Point Spread Function (PSF) of the LAT
to assign to each event its probability of 
coming from the pulsar \citep{ker11}. 
For our timing analysis we used only barycentered events with a probability greater than $0.01$.
The rotational ephemeris used in \citet{abd13} spans only three years:
we extended it, using a weighted Markov Chain Monte Carlo algorithm \citep[MCMC, see e.g.][]{wan13}.
Adding 6 months of data in each iteration, we re-evaluated the timing solution using the H-test \citep[see e.g. ][]{dej10}.
Apart from the glitch reported in \citet{ray11}, we detected a second one (at MJD=56290, see Table~\ref{tab1}). By analyzing separately
the three time intervals (before the first glitch, between the two and after the last one)
we obtained the best ephemeris and light curve for each period.
Then, following \citet{abd13} we fitted each curve with a composite model
encompassing a constant and three gaussians. Using the relative positions of each gaussian maximum, we extracted
the relative phases of the three light curves and built the 5-years J1813 ephemeris, reported in
Table~\ref{tab1}.

\begin{figure*}
\centering
\includegraphics[angle=0,width=15cm]{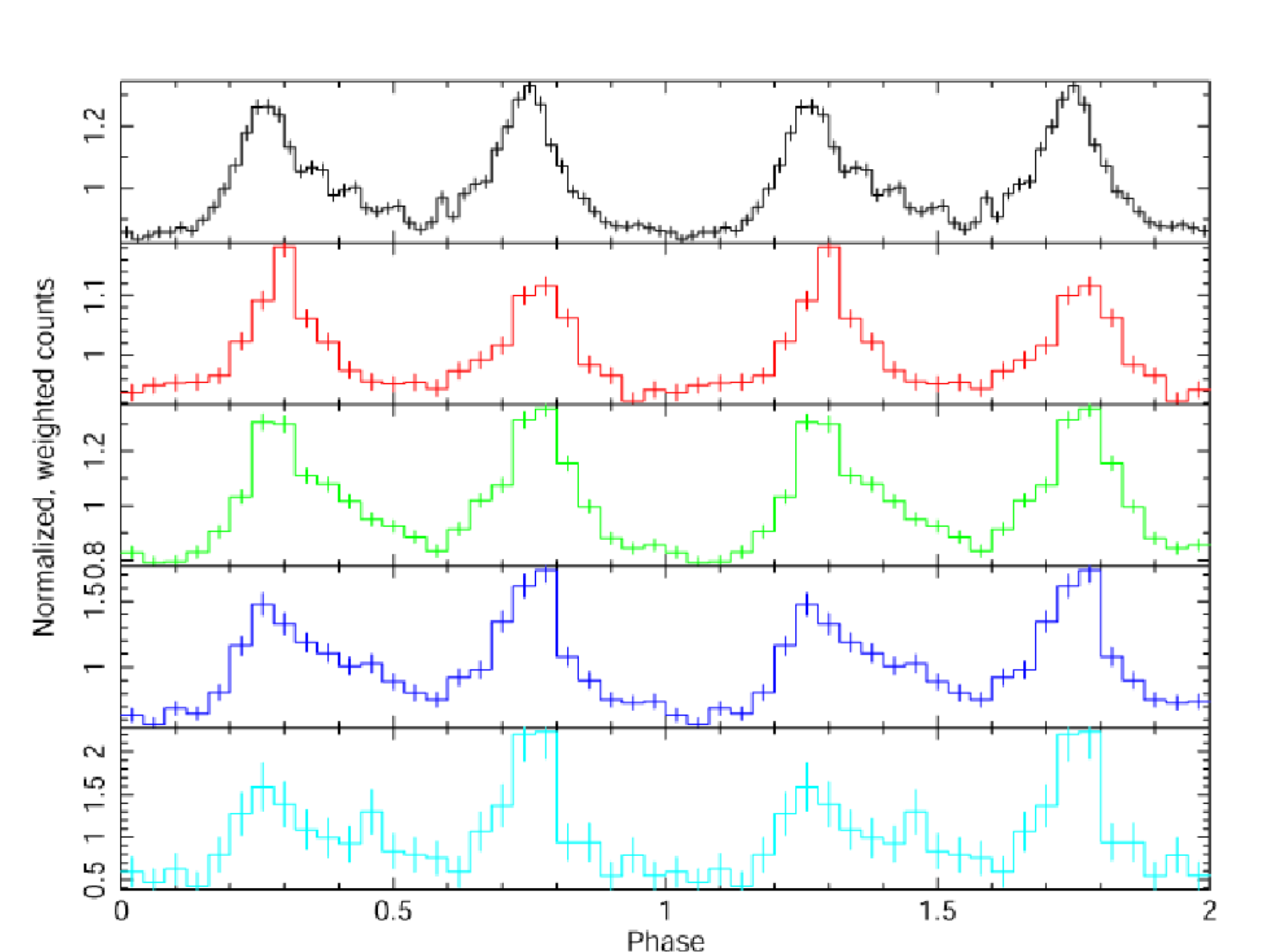}
\protect\caption{{\footnotesize Normalized, weighted {\it Fermi} $\gamma$-ray light curves, using the ephemeris described in Section~\ref{gray} and with MJD$_0$=56362.0. From the top panel, the curves are in the: $>$0.1 GeV, 0.1-0.3 GeV, 0.3-1 GeV, 1-3 GeV, $>$3 GeV energy ranges, respectively. The curves have been renormalized by dividing each bin by N$_{counts}$/N$_{bins}$, where N$_{counts}$ is the 
total weighted number of events in the energy range and N$_{bins}$ the number of bins. 1$\sigma$ errors are shown.}
\label{fig-glc}}
\end{figure*}

Using our ephemeris we assigned a rotational phase to each $\gamma$-ray event and filled a 100-bin $>$0.1 GeV phase histogram, 
with bin uncertainties taking into account the photon weights (see Figure~\ref{fig-glc}).
Being J1813 a radio-quiet pulsar, phase 0 was chosen arbitrarily at MJD$_0$=56362.0.
Our light curve is consistent with the one from \citet{abd13}
and is characterized by two peaks: the maximum of the first peak is at phase
0.258$\pm$0.003 and the second at phase 0.743$\pm$0.002.
The separation between the peaks is 
0.485$\pm$0.003 in phase, with a bridge of emission between the peaks.
While the normalization of the two peaks is similar, the first one is asymmetric, with a clear trail.
Following the prescriptions of \citet{abd13}, we obtained
an acceptable fit ($\chi^2_{red}$=1.78, 40dof, null hypothesis probability, nhp=0.002) using two gaussians to
describe the first peak and one for the second.\footnote{The nhp is the probability of obtaining a test statistic result at least as extreme as the one that was actually observed, assuming that the null hypothesis is true.}

To perform a $>$0.1 GeV $\gamma$-ray phase-resolved spectral analysis on the 5-year dataset, we rebinned the light curve into 25 bins.
For each phase bin, we re-ran the binned likelihood spectral analysis
to search for variations in the spectral parameters as a function of the pulsar phase. We used the same region
of the phase-averaged analysis but we fixed spectral parameters of all the other sources at the
best fitted ones. We left free to vary all the spectral parameters of J1813, as well as the Galactic and isotropic spectral parameters.
Anyway, we note that by freezing the Galactic and isotropic spectral parameters we obtain consistent results.
A simple $\chi^2$ to test the variation of the best fitted parameters,
leaving all the pulsar spectral parameters free to vary, is not adequate since such parameters
are correlated and a single parameter variation cannot describe the overall spectral change.
To search for spectral variation we can compare the best fit Test Statistic (TS) of models with
one or more parameters left free to vary in each bin.
While usually a source TS is used to gauge the source significance against a model that
does not contain such source (model 0) \citep{mat96,cas79}, we can build a TS
that expresses the likelihood ratio between the source spectral model
with a parameter fixed (model 1) and the same model with the same parameter left free (model 2) by using:

\begin{dmath}
TS_{2vs1} = -2 ln\frac{L_1}{L_2} = -2 ln\frac{L_0}{L_2} + 2 ln\frac{L_0}{L_1} = TS_{2vs0} - TS_{1vs0}
\end{dmath}

For high statistics, such TS follows a distribution similar to a $\chi^2$ with one degree of freedom.
All the boundary conditions, as defined in \citet{cas79,pro02}, are verified in this case.
For a phase-resolved spectral analysis, we are considering $N$ (in our case 25) bins. In such a case we can obtain
a TS$_{2vs1,tot}$ that expresses the probability that a given parameter is constant during the phase as:

\begin{dmath}
TS_{2vs1,tot} = -2 ln\prod_{i=1}^N \left(\frac{L_1}{L_2}\right)_i = \sum_{i=1}^N \left(-2 ln\frac{L_{0,i}}{L_{2,i}} + 2 ln\frac{L_{0,i}}{L_{1,i}}\right) = \sum_{i=1}^N (TS_{2vs0,i} - TS_{1vs0,i})
\end{dmath}

\begin{figure*}
\centering
\includegraphics[angle=0,width=15cm]{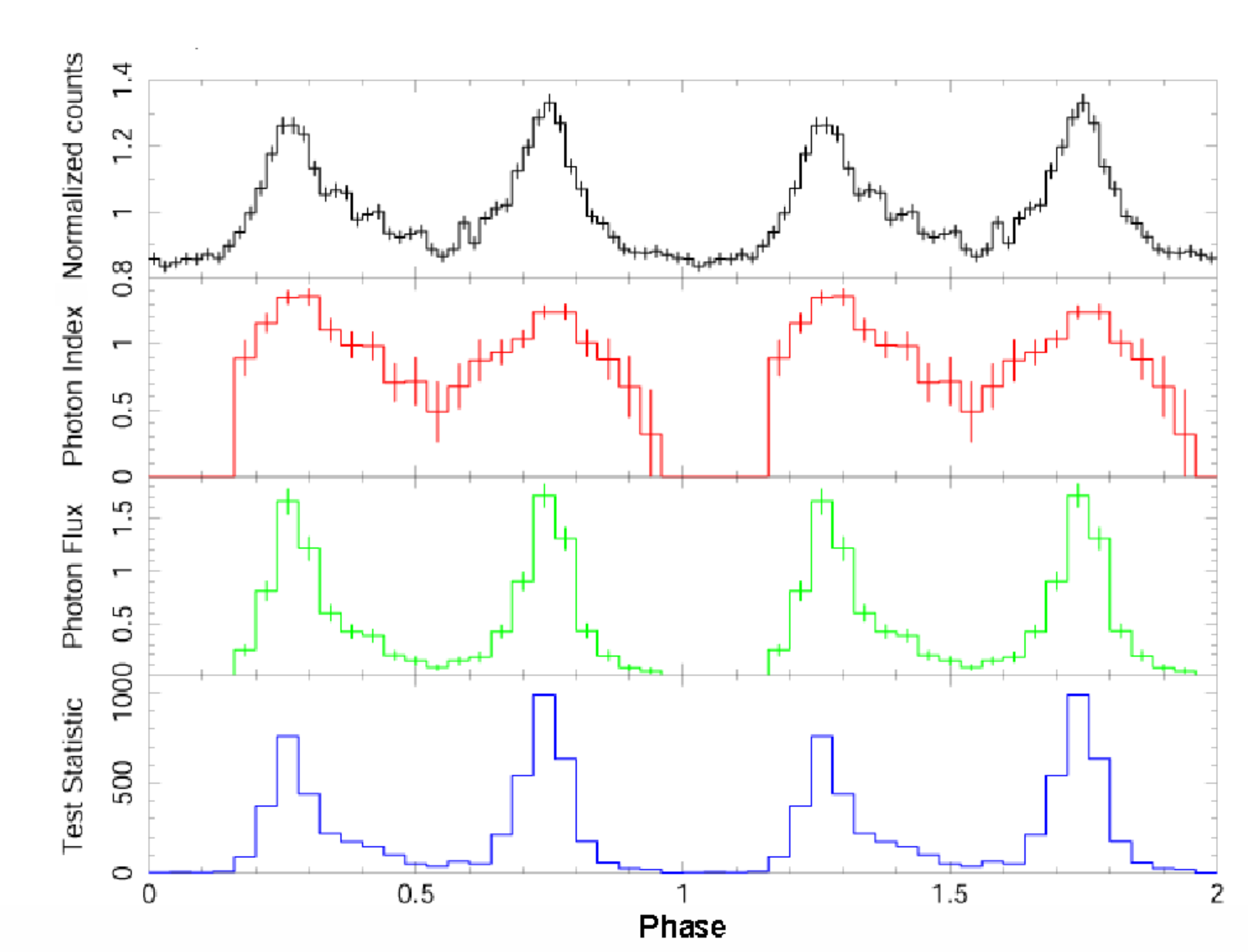}
\protect\caption{{\footnotesize Results of {\it Fermi} phase-resolved spectroscopy leaving free to vary only normalization and photon index. Black: $>$0.1 GeV normalized, weighted light curve; red: best fitted photon indexes (only for bins with TS$>$25); green: best fitted photon fluxes (10$^{-8}$ photons cm$^{-2}$ s$^{-1}$); blue: Test Statistic of the fits. 1$\sigma$ errors are shown.}
\label{fig-pi}}
\end{figure*}

For high statistics, such TS follows a distribution similar to a $\chi^2$ with $N$ degrees of freedom.
Applying this to our phase-resolved spectroscopy,
the TS of variable normalization+index and variable normalization+cutoff, both compared to the only-normalization variation,
are TS$_{2vs1,tot}$=1780 and TS$_{2vs1,tot}$=624, respectively.
We can therefore conclude
that there is a spectral variation of J1813 with phase. The change of the photon index is much more compelling
than the variation in the cutoff energy.
As apparent in Figure~\ref{fig-pi}, the spectrum softens during each peak,
while it is harder during the bridge between the two peaks.\\
At variance with the finding of \citet{abd13} our 5-year light curve does not show a significant off-pulse emission.
As off-pulse interval we chose the bins in which the source has TS$<$25 both in each bin and in the entire interval.
Indeed, when selecting the phase interval 0.96-0.16 the source is barely detected at the $\sim3\sigma$ level.
We note that, due to the improvement in the statistics and models, we the off-pulse interval we use is shorter than the one in 2PC.
Improved models of diffuse Galactic and isotropic background emission are probably responsible for this result.

\section{X-ray observations and analysis}

\begin{figure*}
\centering
\includegraphics[angle=0,width=15cm]{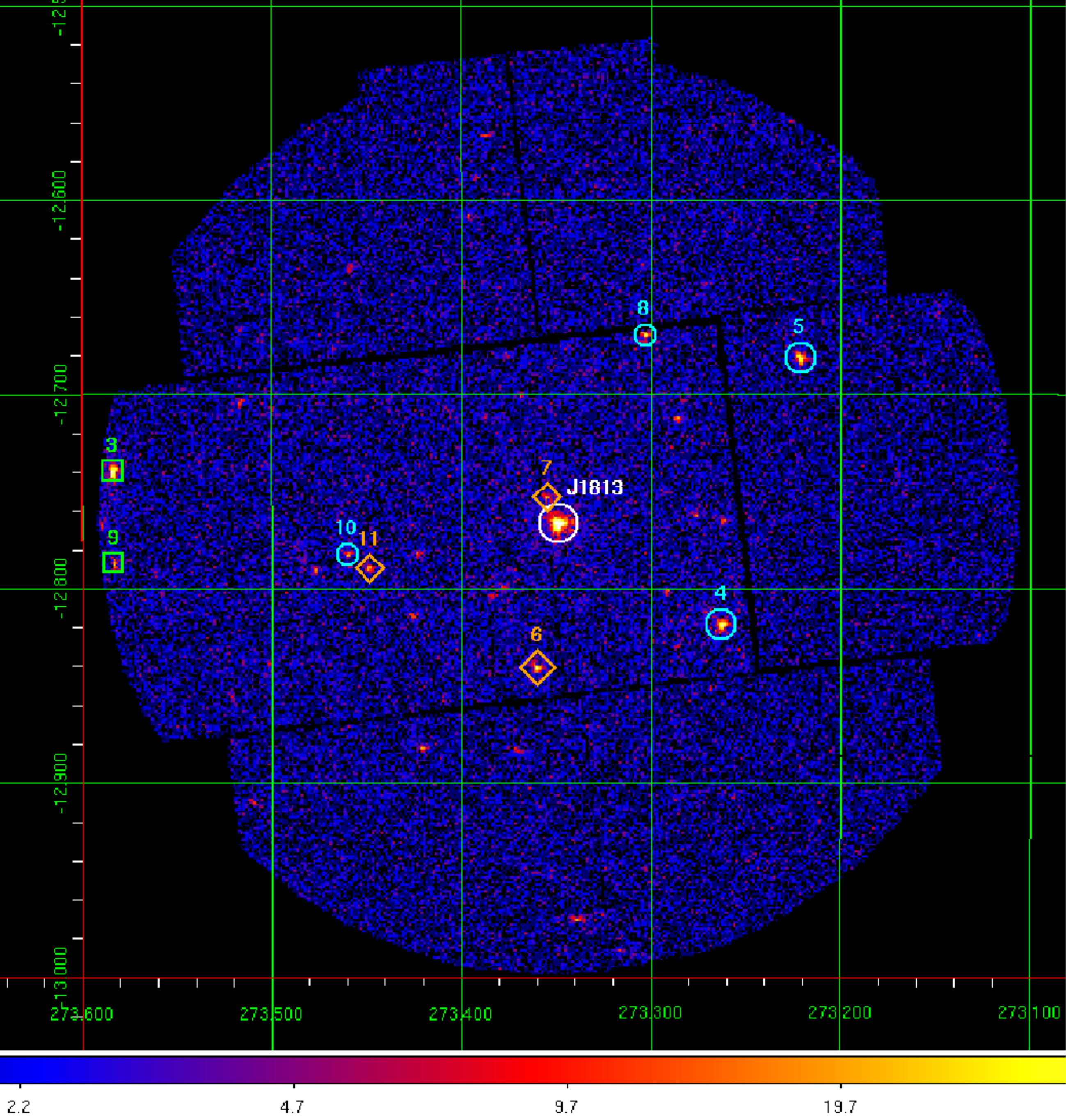}
\protect\caption{{\footnotesize 0.3-10 keV Field of View of {\it XMM-Newton} MOS2 camera. The pulsar is circled in white,
the X-ray emitting stars in orange, the AGN in cyan and the unidentified sources in green.}
\label{fig-fov}}
\end{figure*}

Figure~\ref{fig-fov} shows the 0.3-10 keV {\it XMM-Newton} FOV.
Source detection using maximum likelihood fitting was done
simultaneously on each of the EPIC-PN, MOS1, and MOS2 with the SAS
tool {\tt edetect\_chain}. 
We also performed a source detection on the {\it Chandra} dataset
by using the CIAO tool {\tt wavdetect}.

\begin{figure*}
\centering
\includegraphics[angle=0,width=15cm]{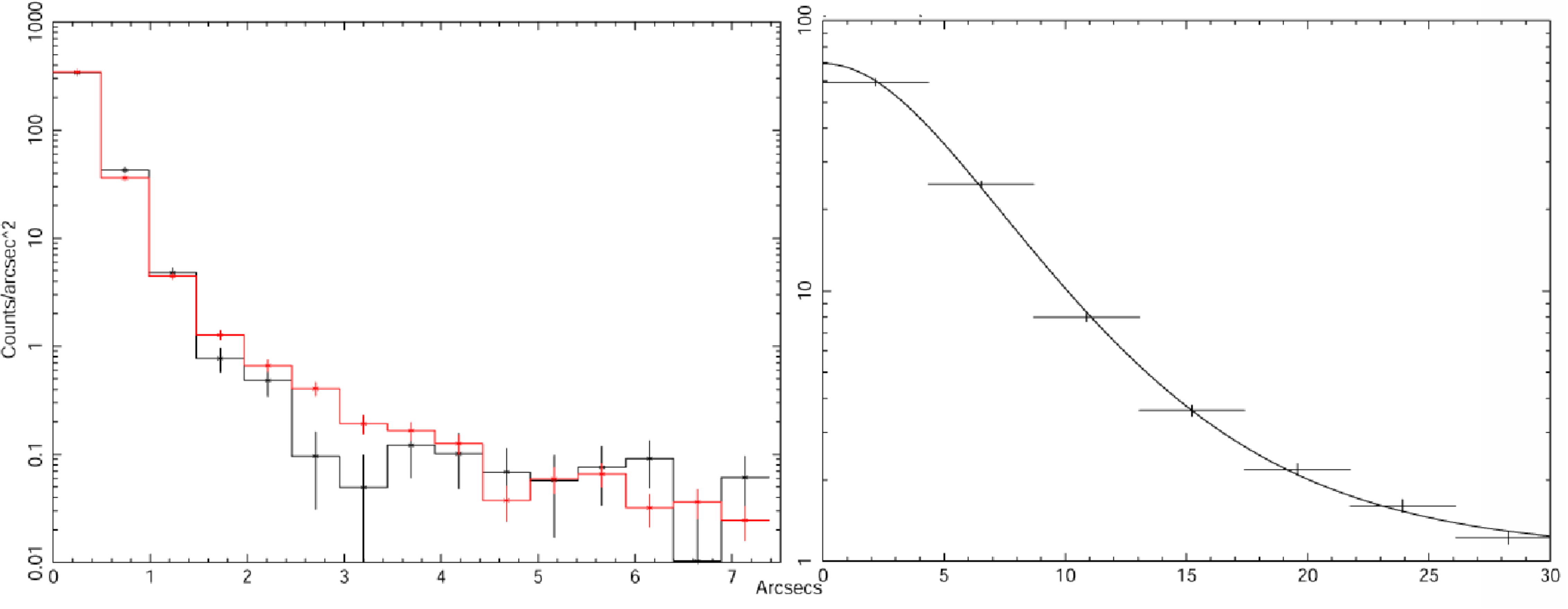}
\protect\caption{{\footnotesize {\it Left}: The {\it Chandra} radial brightness profile of J1813 (black) and the simulated one (red). {\it Right}:
The {\it XMM-Newton} radial brightness profile of J1813 and its best fit with a constant plus King function.
Both the fits point to lack of diffuse emission down to fractions of arcsec.}
\label{fig-psf}}
\end{figure*}

The best X-ray position of the pulsar is 18$^h$13$^m$23$^s$.77, -12$^{\circ}$45$'$59.9$''$ (0.015$''$+0.6$''$ 90\% statistical plus systematic errors).
We analyzed the pulsar radial brightness profiles in {\it XMM-Newton} and {\it Chandra} datasets and compared them with
the theoretical PSFs.
The PSF of the EPIC-PN camera onboard {\it XMM-Newton}
is best described by an off-axis, energy-dependent King function \citep{rea04}.
The full width half maximum of the PSF for an on-axis source at 1.5 keV is typically less than 12.5$''$
for the PN camera and 4.4$''$ for the two MOS detectors\footnote{\url{http://xmm.esac.esa.int/external/xmm\_user\_support/documentation/sas\_usg/USG/}}.
The theoretical {\it Chandra} PSF is much more complicated and largely off-axis dependent;
its evaluation requires simulations of the specific observation using {\tt Chart} and {\tt MARX}.
The observed PSFs, the fit with the {\it XMM-Newton} theoretical one and the {\it Chandra} PSF simulation for a
point-like source are shown in Figure~\ref{fig-psf}.
The observed {\it XMM-Newton} brightness profile is well fitted by the theoretical PSF ($\chi^2_{red}$=1.1 , dof=5, nhp=0.39)
and the observed and simulated {\it Chandra} profiles agree (fitting the residuals with a constant we obtain
$\chi^2_{red}$=2.4, dof=15, nhp=0.01). We therefore conclude that no extended emission is detected down to a fraction
of an arc second. The {\it Suzaku} detection of a nebula, reported in \citet{abd13}, 
is due to source \#7, located 50$''$ from the pulsar.

Figure~\ref{fig-fov} shows the brightest sources in the PN field of view.
The study of the line-of-sight absorption of such sources 
could allow us to constrain the
pulsar distance. Indeed, after selecting candidate Active Galactic Nuclei 
(AGN) in the FOV, it is possible to measure from their spectra
the total Galactic column density 
in the direction of J1813. Next, using the pulsar column density, we can get 
an estimate of its distance with respect to the
edge of the Galaxy. Such estimate could be refined if bright X-ray 
and optical stars (with known distance)
were also present
\citep[see e.g.][]{mar13,mar14}.

Based on the study of spectra and possible optical counterparts,
we classified serendipitous sources as AGNs or candidate stars (see Appendix). 
Our exercise allowed us to identify four very likely AGN and three stars. 

\begin{figure*}
\centering
\includegraphics[angle=0,width=15cm]{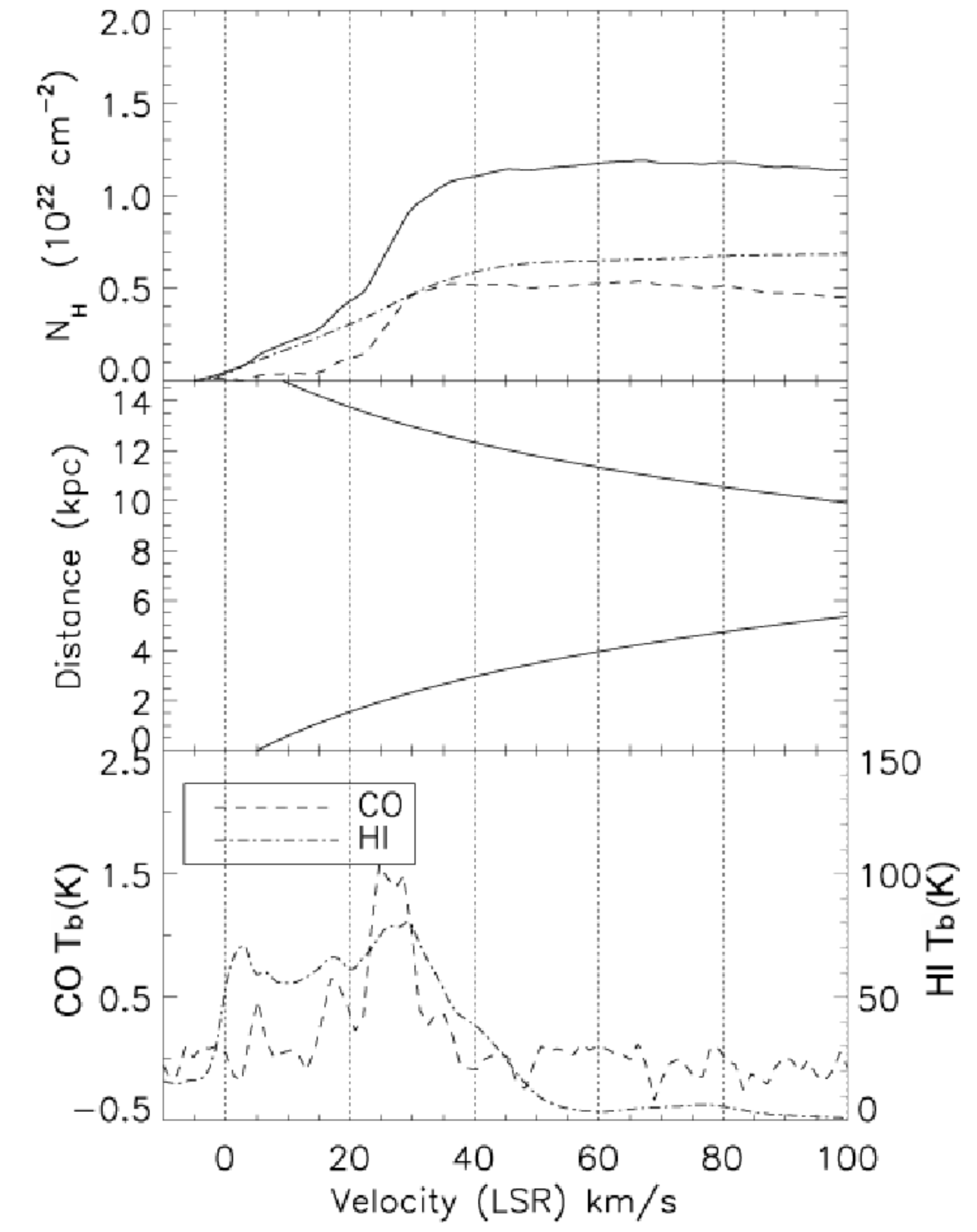}
\protect\caption{{\footnotesize {\it Top:} cumulative absorption column N$_{\rm H}$ (solid line) towards J1813 derived
from atomic (HI, dotted line) and molecular ( $^{12}$CO, dashed line) gas. {\it Middle:}
Distance as a function of radial velocity derived from the Galactic rotation curve
model of \citet{hou09}. {\it Bottom:}  $^{12}$CO  (dashed line) and HI (dotted line)
spectra at the position of the pulsar.}
\label{fig-molh}}
\end{figure*}

The spectra of the AGN show very high values of column density ((1 - 2)$\times10^{22}$ cm$^{-2}$),
higher than the value of $7 \times$ 10$^{21}$ cm$^{-2}$ obtained from the 21 cm HI sky survey of \citet{kal05}.
Given the unexpectedly high values of column density of the sources inside the {\it XMM-Newton} FOV,
we searched for the presence of molecular clouds in that region.
The all-sky model of dust emission from {\it Planck} \citep{abe14} allow us to estimate the
dust temperature uniformly over the whole sky, providing an improved estimate of the
dust optical depth compared to previous all-sky dust model. The region of J1813 is characterized
by a higher temperature than the mean of that latitude, pointing to an absorption higher than usual.
\citet{dob11} presents an atlas and catalog of dark clouds derived from the
2 Micron All Sky Survey Point Source Catalog \citep[2MASS PSC,][]{skr06}
and reports four dark clouds within the {\it XMM-Newton} FOV, with a structured pattern.
The discrepancy between the Galactic absorption and the $N_H$ values of our AGN-like sources should be
ascribed to the presence of such an irregular pattern of dark clouds.\\
The best-fit $N_H$ value of J1813 will indicate if the pulsar is located in front or in the rear of those clouds.
With the distance to the clouds known, this will become an important estimator of the pulsar distance.\\
In fact, a comparison of the X-ray absorption column along the line of sight obtained with the 
column density  derived from the atomic (HI) and molecular ($^{12}$CO,
J=1$\rightarrow$0 transition line) gas can be used to provide a lower limit on the distance of J1813.
The data from the $^{12}$CO \citet{dam01} CfA survey and from the HI Parkes Galactic
all-sky survey \citep{mcc09,kal05} are used.
The CO-to-H$_{\rm 2}$ mass conversion factor used is 1.8$\times10^{20}$
cm$^{-2}\,$K$^{-1}\,$km$^{-1}\,$s \citep{dam01} and the HI brightness temperature to column density is
$1.82\times10^{18}\,$cm$^{-2}\,$K$^{-1}\,$km$^{-1}\,$s \citep{dic90}.
The Galactic rotation curve model of \citet{hou09} is used to translate the measured velocities into distances.
All absorbing material is assumed to be at the near distance allowed by the Galactic rotation curve.
As shown in Figure~\ref{fig-molh}, the main $^{12}$CO absorption feature along the line of sight
is located at a radial velocity relative to the local standard of rest (LSR) of $V_{\rm LSR}=27$ km s$^{-1}$ 
corresponding to an integrated column density of 1$\times10^{22}$ cm$^{-2}$ (HI+$^{12}$CO).
Therefore if the fitted X-ray column density of J1813 is higher than this value
(see next Section), we conclude that the pulsar is located behind the clouds at a
distance $>$ 2.5 kpc.   

\subsection{X-ray Spectral Analysis} \label{xspec}

To study the spectrum of J1813, we simultaneously fitted spectra from {\it XMM-Newton}, {\it Chandra}
and {\it Suzaku}. For {\it XMM-Newton} we chose an extraction radius
of 25$''$, in order to avoid contamination from the bright source
located at 50$''$. We obtained 6072, 2581 and 2757 net counts in the 0.4-10 keV energy range in
the PN and the two MOS detectors respectively, taking into account the background contribution (9\%, 4\% and 4\% respectively). 
For {\it Chandra} we chose an extraction radius of 2$''$ and we obtained 1494 net counts
in the 0.3-10 keV energy range, taking into account the background contribution (less than 0.1\%).
For {\it Suzaku} we chose an extraction radius of 70$''$ to minimize the contamination from the nearby source.
With the chosen extraction radius such a contamination is expected to be negligible ($\sim$1.5\% of the total counts).
From {\it Suzaku} we obtained 403, 318 and 435 net counts in XIS 0,1 and 3 respectively,
taking into account the background contribution (16\%, 37\% and 13\% respectively). 

The very hard spectrum of J1813 is well fitted ($\chi^2_{red}$=1.09, null hypotesis probability=0.08) by a power law
with $\Gamma$=0.85$\pm$0.03, absorbed by a column density N$_H$=1.56$\pm$0.07 $\times10^{22}$ cm$^{-2}$ (1$\sigma$ confidence level).
A composite thermal plus non-thermal model is not statistically needed. In fact,
an $F$ test \citep{bev69} shows that the probability for a chance improvement by using the composite spectral model is 0.003, less than
a 3$\sigma$ significance level. The unabsorbed 0.3-10 keV flux of J1813 is 1.08$\pm$0.01 $\times10^{-12}$ erg cm$^{-2}$ s$^{-1}$,
leading to a $\gamma$-to-X flux ratio of 234$\pm$6, three times less than the lowest one of the radio-quiet pulsar family \citep{mar11,abd13}.
Such a low value of the $\gamma$-to-X flux ratio is different from the higher value reported in 2PC for J1813
(1840$_{-610}^{+330}$). That result was based only on the short {\it Suzaku} and {\it Swift}
observations, thus the low statistic prevented a correct characterization
of the source. Moreover, the extracted {\it Suzaku} spectrum was contaminated by
the nearby star (source \#7 in Figure~\ref{fig-fov}): such a soft spectrum prevented
them from a correct evaluation of the column density, resulting in a lower
value of the unabsorbed X-ray spectrum on which the $\gamma$-to-X flux ratio is based. We
also note that in 2PC 30\% of J1813 total flux was expected to come from
thermal and nebular emission.

\subsection{X-ray Timing Analysis} \label{xtim}

To search for X-ray pulsations from J1813, we used the SAS tool {\tt
  barycen} to barycenter the PN events using the precise {\it Chandra} pulsar position.
In order to improve the sensitivity to pulsations, we decided to apply a
photon weighting technique similar to the one used for {\it
  Fermi}-LAT, assigning to each photon a probability of coming from
the pulsar, in order to help with background rejection and improve the sensitivity to pulsations.
\citet{ker11} notes that this technique is applicable to any photon-counting instrument in which sources are not
perfectly separated from their background, e.g., searches for
X-ray pulsation in observations of a pulsar. While in the X-ray domain
the positional errors are much smaller than in the $\gamma$-ray band,
the problem of superposition of sources is more critical. Indeed, the wide
range of spectral shapes for different source classes in X-rays and the complexity of the
background of X-ray telescopes \citep[for {\it XMM-Newton} see e.g. ][]{kun08}
make such techniques as important as in the $\gamma$-ray domain.\\
To this end, we developed and used a Python tool to assign to each photon weights quantifying the probability
that such an event comes from each of the sources within the region of interest. The tool requires:\\
- the position, (best fitted) spectral model, flux and fitted PSF of each source;\\
- the (best fitted) spectral model and flux of the background.\\
The tool produces columns of weights that are added to the events file, replicating
the {\it Fermi} tool {\tt gtsrcprob}.
Tests conducted on a sample of pulsars \citep[e.g. the magnificent seven pulsars and Geminga,][]{tre01,car04}
resulted in a significant improvement of the H-value with respect to unweighted periodicity tests with optimized spatial and
energy cuts. The ratio $H_{weight}/H_{unweight}$ ranges from 1.2 for bright, soft sources
to 2 for faint sources with hard spectra.

Similarly to the $\gamma$-ray timing analysis, for J1813 we used a weighted MCMC
algorithm (20 harmonics) to search for the best pulsar period during the 1-day long {\it XMM-Newton} observation, also testing
the extended {\it Fermi}-LAT ephemeris we found.
The best frequency at MJD$_0$ is 20.80107408901 Hz (H-value=12092, where an H-value of 95 yields a 5$\sigma$ significance),
consistent with the one from {\it Fermi}-LAT ephemeris (that yields H-value=12088).
For comparison, an unweighted test with the best energy and spatial cuts yields an H-value of 11123.
Such best fit H-values have been obtained by using not-randomized {\it XMM-Newton} events.
In the Small Window mode of the EPIC-PN camera, arriving photons are read only during a cycle of 3.9809 ms (=integration time),
then the charges are transferred (transfer time=0.068 ms) and read (readout time=1.521 ms). The real arrival time of each photon is
then stored as a multiple of the frame time (=integration+transfer+readout time) of $\sim$5.7 ms
\citep{kus99}.
For standard analysis, a tool included in the SAS {\tt epproc} usually randomizes the arrival time of each event within the 5.7 ms windows.
Here, we chose not to randomize the arrival times in order to achieve a better timing resolution for our fast rotating pulsar.
The un-randomized event file basically consists of sets of photons with the same arrival times, at the middle of each
5.7 ms window. By using this type of events, we avoid the error from the randomization process on the entire 5.7 ms window
(while the integration time is only 4 ms).
For comparison, a randomized, unweighted test on J1813 with the best energy and spatial cuts gives an H-value of 8735.

\begin{figure*}
\centering
\includegraphics[angle=0,width=15cm]{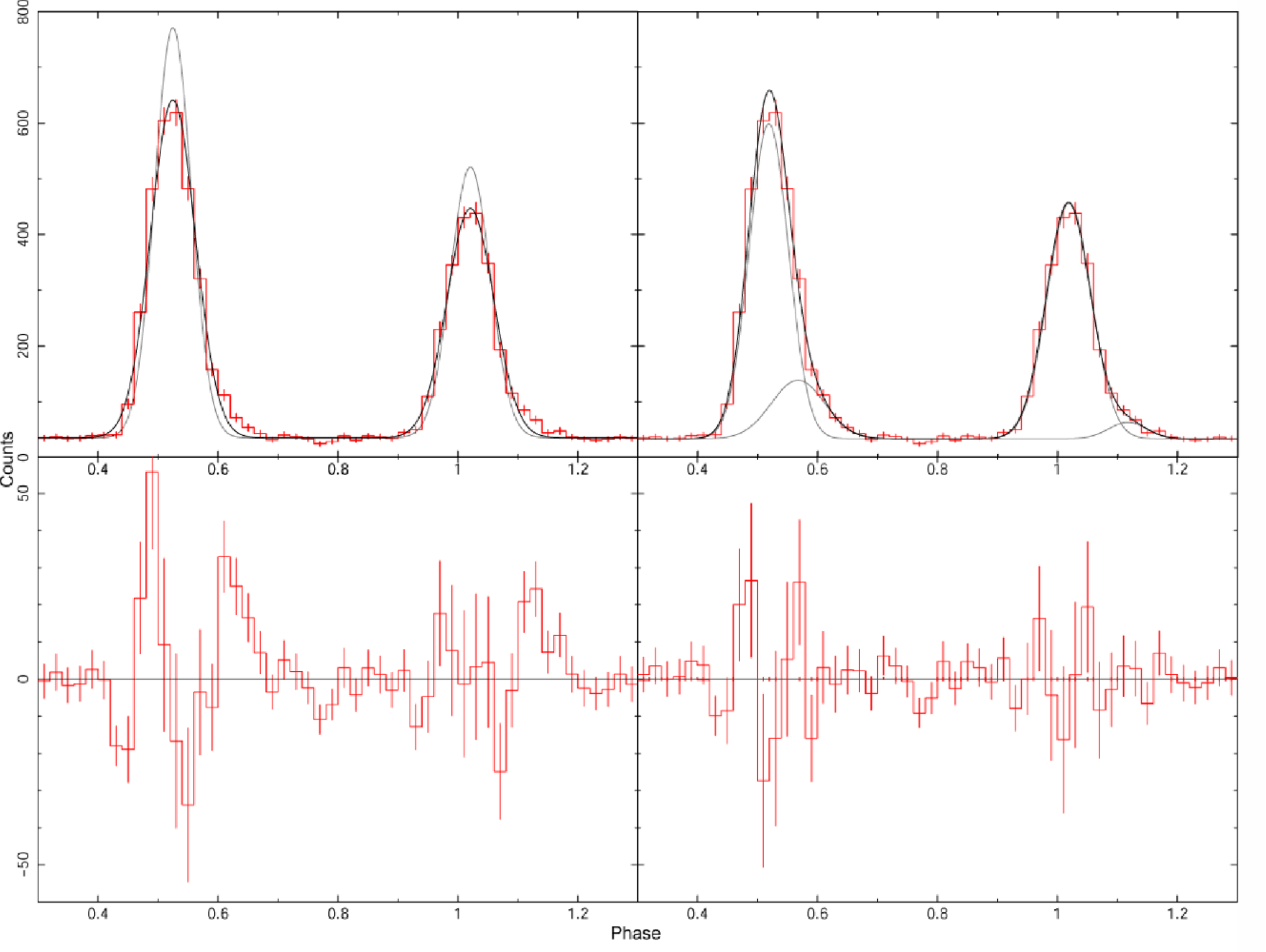}
\protect\caption{{\footnotesize The 0.3-10 keV weighted, non-randomized light curve of J1813 is shown in red.
{\it Upper-left}: Model of the X-ray light curve before
the distorsion due to the {\it XMM-Newton} frame time (grey) and after the simulation (black), that is also the best fit of
the light curve by using 2 gaussians and a constant. {\it Lower-left}: residuals. {\it Upper-right}: Best fit of the X-ray light
curve (black) using 4 gaussians and a constant; in grey the single gaussians components are shown. {\it Lower-right}: residuals.}
\label{fig-gaux}}
\end{figure*}

\begin{figure*}
\centering
\includegraphics[angle=0,width=15cm]{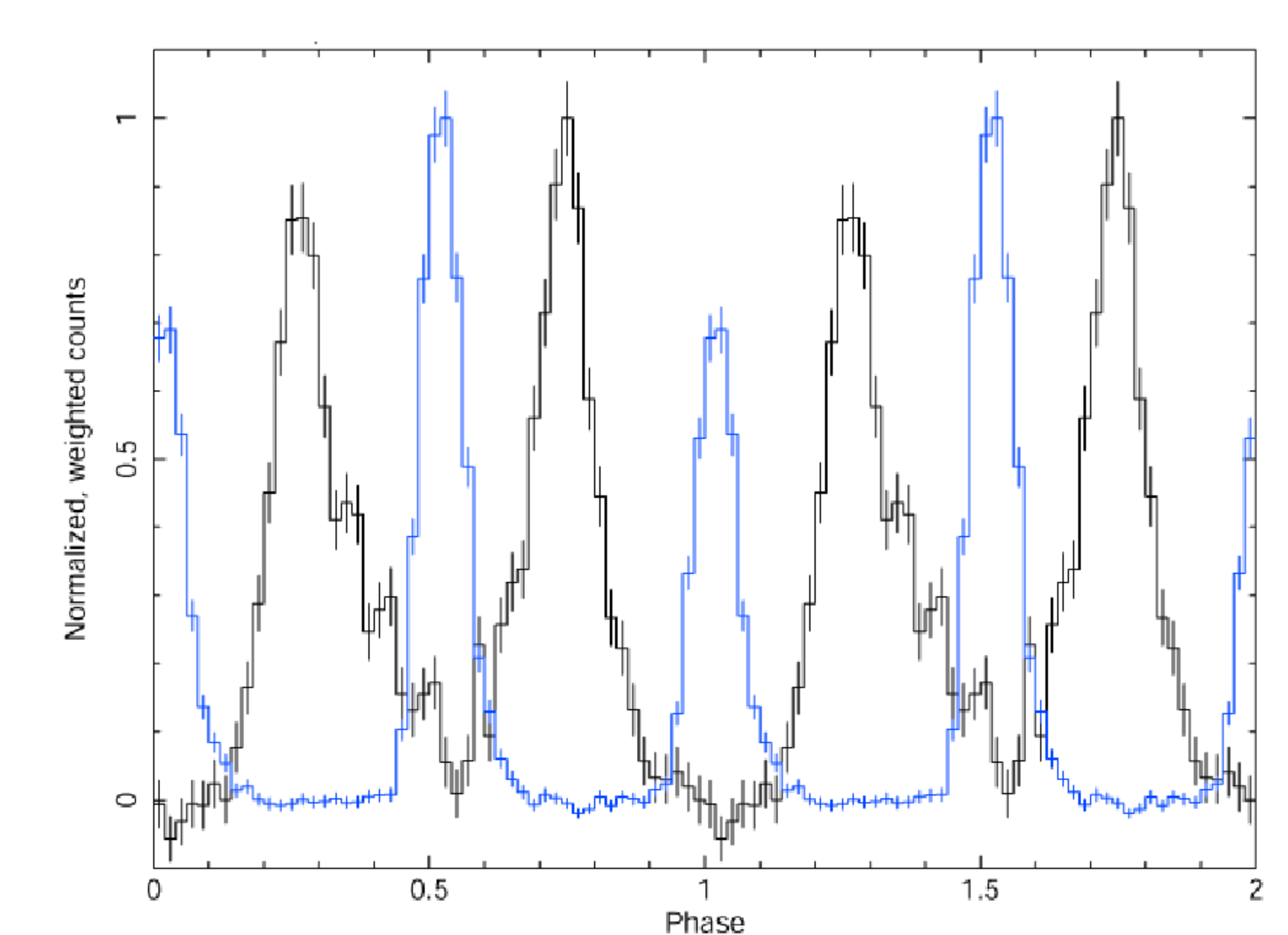}
\protect\caption{{\footnotesize Phased $>$0.1 GeV {\it Fermi} light curve (black)
and 0.3-10 keV {\it XMM-Newton} light curve of J1813 (cyan), in phase.
1$\sigma$ errors are reported. The normalization is defined as in Figure~\ref{fig-glc}.}
\label{fig-xgamma}}
\end{figure*}

The resulting X-ray light curve shows two very sharp peaks, about 8 ms wide, with an off-pulse component
detected with a 17$\sigma$ significance.
Here, we define as off-pulse the sum of phase bins for which the count rate can be fitted by a constant. All the bins
that deviate more than 3$\sigma$ from the fitted value are considered on-pulse.
In such a way, we obtain two off-pulse intervals, between 0.15-0.4 and 0.65-0.9 in phase.
The pulse profile is expected to be heavily affected by the PN camera frame time binning.
Thus we developed a Python script to simulate the deformation of simple input light curves due to
such PN (Small Window mode) readout cycles.\\
Simple input step functions cannot reproduce the measured pulse profile. A two-gaussian
model is instead able to reproduce the observed profile (see Figure~\ref{fig-gaux}).
By using our simulation we concluded that the X-ray pulsar profile (before the deformation due to
{\it XMM-Newton} frame time binning) is well described by two gaussians
located at phases 0.0205$\pm$0.0005 (peak1), and 0.5248$\pm$0.0007 (peak2), with standard deviations of 
(3.30$\pm$0.08)$\times10^{-2}$ and (3.05$\pm$0.05)$\times10^{-2}$;
the normalization of the first peak is a factor 1.48$\pm$0.05 lower than that of the second one.
The separation between the first and the second peak is 0.4957$\pm$0.0009, a value in agreement
(within 3$\sigma$) with the $\gamma$-ray one.
Both X-ray peaks are slightly asymmetric, with tails, and their fitting requires two gaussians,
reminiscent of the first $\gamma$-ray peak.
Using an $F$ test, we determined that the probability for a chance improvement is 4.1$\times10^{-7}$,
pointing to a significant improvement by adding two more gaussians. The peak of the first $\gamma$-ray gaussian
lags the peak of the first X-ray gaussian by 0.237$\pm$0.002 in phase and the peak of the second $\gamma$-ray gaussian
lags the second X-ray peak by 0.218$\pm$0.003 in phase (Figure~\ref{fig-xgamma}).

By definition, the weighted light curve is background subtracted. Here, we define the pulsed fraction as M$_{gau}$/(M$_{gau}$+C),
where M$_{gau}$ is the maximum of the wider gaussian and C the constant in the model.
We considered the best simulated light curve model in order to exclude the {\it XMM-Newton} frame time
binning distorsion. The pulsed fraction of J1813 in the 0.3-10 keV energy range is 96$\pm$3\%. We note that the remaining percentage
represents the maximum allowed count rate of a possible nebula.
No statistically significant variations of the pulsed fraction are measured by using different energy ranges, pointing to
one-component spectral models for J1813.

We performed phase-resolved spectroscopy with different selections of phase bins.
In order to detect any possible spectral variation with phase, we analyzed on and off-pulse spectra
(as defined in Section~\ref{xtim}) as well as the spectra of each peak. We also fitted the first and last half
of each peak. Lastly, we divided in three equal parts each peak.
We fitted together spectra obtained from each of the described divisions of the phase.
In all the cases fixing the photon indexes we found acceptable spectral fits.
An $F$ test shows a non-negligible probability for a chance improvement by freeing the photon index
(in each case $>$1.5$\times10^{-3}$). We conclude that no spectral variation is seen in the X-ray band as a function of the pulsar phase, with a
3$\sigma$ upper limit of 0.08 in the photon index variation between on and off-pulse phases.

\section{The hard X-ray band and Spectral Energy Distribution}

Searching at the position of J1813 in the hard X-ray band ($\sim$100 keV),
we found a hint of a detection with {\it INTEGRAL} IBIS/ISGRI \citep{leb03}.
We used the automated HEAVENS online tool \footnote{\url{http://www.usdc.unige.ch/heavens/}}
to create a counts map, sensitivity map and light curve of the source with all the public
{\it INTEGRAL} observations.
The possible steady source has a count rate of 0.12$\pm$0.02 counts s$^{-1}$ in the
30-520 keV energy band, that corresponds to a flux of $\sim5.3$ $\times$ 10$^{-11}$erg cm$^{-2}$ s$^{-1}$.
Although the significance of the detection is $>6\sigma$, such a source comes from an automated
script instead of from a dedicated analysis. We therefore conservatively decided to treat it as an upper limit. 

\begin{figure*}
\centering
\includegraphics[angle=0,width=15cm]{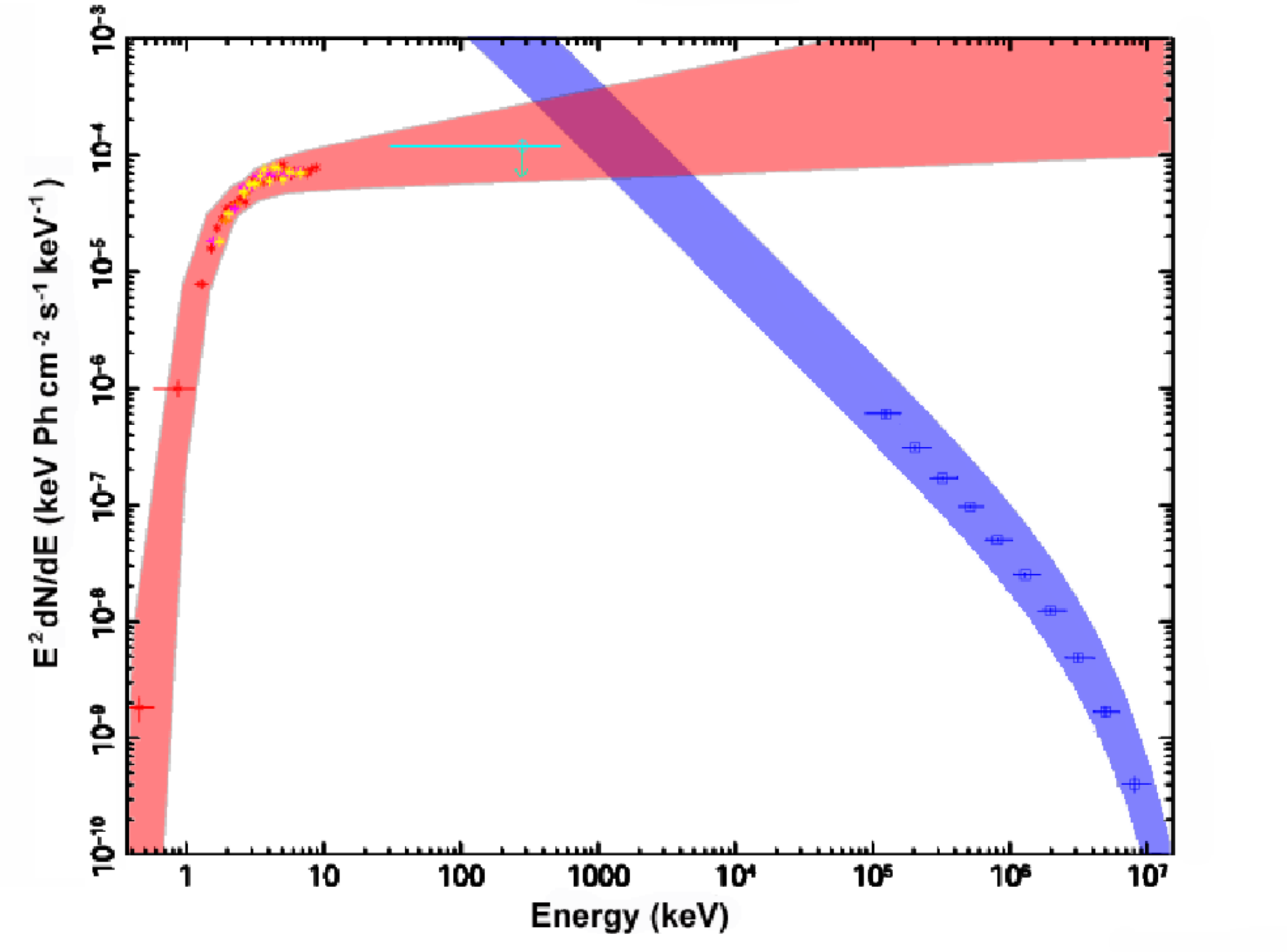}
\protect\caption{{\footnotesize Spectral energy distribution of J1813. Red, magenta, orange and yellow star points mark the
unfolded 0.4-10 keV spectra from {\it XMM-Newton} PN, MOS1, MOS2 and {\it Chandra}, respectively. Blue square points
result from binned likelihood spectral analysis of {\it Fermi} data of logarithmically uniform energy bins.
The cyan round point reports the hint of detection with {\it INTEGRAL} IBIS/ISGRI, that we
conservatively chose to treat as an upper limit. The red area is the X-ray 1$\sigma$ butterfly for the absorbed model:
any X-ray-band absorbed power-law model that is drawn on the plot which is not fully
contained in the envelope is excluded by the data at the 1$\sigma$ confidence level.
Similarly, the blue region is the $\gamma$-ray 1$\sigma$ butterfly, using a power-law with exponential cutoff.
We note that these regions are verified only in the X-ray (0.3-10 keV) and $\gamma$-ray (0.1-100 GeV) band and then estrapolated
to the full plot.}
\label{fig-sed}}
\end{figure*}

In order to find the spectral energy distribution (SED) points in the X-ray band we regrouped the {\it XMM-Newton} and {\it Chandra} spectra.
We plotted the X-ray unfolded spectrum with XSPEC by using the best fitted spectral model reported in Section~\ref{xspec}.
To find the SED points in the $\gamma$-ray band, we divided our dataset in logarithmically uniform energy bins.
Then, for each bin we re-ran the binned analysis reported in Section~\ref{gray}.
We derived the 1$\sigma$ confidence $N$-dimensional ellipsoids (where $N$ is the number
of free parameters in the models) from the covariance matrices obtained as an output of XSPEC and {\tt gtlike} for the
X-ray and $\gamma$-ray band, respectively. Then, we simulated 10$^4$ spectra for each band with parameters following the contours
and reported these in Figure~\ref{fig-sed} (butterfly plot). The meaning of the butterfly can be understood as follows:
any absorbed power-law model (for X-rays) or power-law with exponential cutoff (for $\gamma$-rays) that is drawn on
the plot which is not fully contained in the envelope is outside the 1$\sigma$ confidence region for such models,
and hence is excluded by the data at the 1$\sigma$ confidence level.

Figure~\ref{fig-sed} clearly shows the presence of a maximum in the SED in the hard-X domain - $\sim$30-500 keV.
In that band a change in the photon index - sudden or gradual - is apparent.

\section{Discussion}

Although the observed $\gamma$-ray light curve and spectrum of J1813 
is quite typical of $\gamma$-ray pulsars, the X-ray light
curve and spectrum is very atypical.  In other {\it Fermi}-LAT pulsars
with non-thermal X-ray emission, like the Crab or the millisecond pulsar
J1939+2134, the X-ray peaks are in phase or nearly in phase with the
$\gamma$-ray peaks.  This is expected in high-altitude emission models
such as the outer gap \citep{tak07} or slot gap \citep{har08},
where particle acceleration and emission along the last
open trailing field lines up to near the light cylinder produces
all outgoing photons at the same phase for an inertial observer
\citep{rom95,dyk04}.  Such caustic emission
will produce the non-thermal $\gamma$-ray, X-ray and optical pulses at
the same phase in the light curve.  In the outer gap model, the two
non-thermal peaks come from the same pole, but at very different
altitudes. In the slot gap model, the two non-thermal peaks come from
trailing field lines from opposite poles, with the emission along
leading edge field lines producing lower-level off-peak emission.  The
fact that the observed non-thermal X-ray peaks in J1813 are not only
not in phase with the $\gamma$-ray peaks, but are both out of phase by
about one quarter of a period, is not in agreement with both X-ray and
$\gamma$-ray emission being outgoing caustic emission from 
the outer magnetosphere.

Estimates of the emission geometry of J1813 have been obtained from
fitting its light curve with a version of the outer gap and slot gap
models \citep{pie12,pie14}. These fits give an inclination angle
$\alpha = 40^\circ$ and observer angle to the spin axis $\zeta =
87^\circ$ for the slot gap, and $\alpha = 8^\circ$ and  $\zeta =
78^\circ$ for the outer gap model.  As expected, our viewing angle is
large and nearly orthogonal to the spin axis. However, the inclination
angles are at least $40^\circ$ different from the observer angle, as
expected for a radio-quiet pulsar, given that a small $\beta = \alpha
- \zeta$ is required in order to miss the radio beam along the magnetic
pole.  The slot gap is slightly favored over the outer gap in the fit,
but not significantly so. 

In many outer magnetosphere emission models, such as outer gap and
slot gap,  the non-thermal optical to X-ray emission is synchrotron
radiation from electron-positron pairs.  In outer gap models, pairs
are accelerated and radiate in both outward and inward directions, so
it might be possible to see a synchrotron component from inward-going
electrons or positrons. But then it is not clear why the outward-going
radiation is also not visible. In slot gap models, all emission is
assumed to be outgoing since the primary particles are only
accelerated outward and the electron-positron pairs from polar cap
cascades, which radiated the non-thermal synchrotron emission, are also
only outgoing. However, from simulations of global magnetospheres
\citep[see e.g. ][]{spi06,tim06}, currents appear in both
directions since there must be a return current. The actual
composition of these currents is not presently known (the models only
give the macroscopic current density), but it is possible that the
main current consists of electrons flowing out and the return current
of electrons flowing inward.  Recent dissipative pulsar magnetosphere
models \citep{kal12,li12}, derive the
electric field parallel to the magnetic field ($E_{\parallel}$) as
well as the (macroscopic) currents and charge densities.  The
$E_{\parallel}$ components appear in both directions, on different
field lines, so depending on the sign of charge that is present along
each field line, charges could be accelerated and radiate inward on some 
field lines (or even in both directions).  

We have tested the possibility that the X-ray
emission in J1813 is not outgoing but ingoing radiation.  In this
case, geometrically, inward-going emission from leading file lines may
produce caustics that would be out of phase with the outward-going
caustic from trailing edge field lines.  We simulated ingoing emission radiated tangent 
to field lines using a geometric representation of the slot gap, known as two-pole caustic 
geometry \citep{dyk03}.  Uniform emissivity was assumed, along the field lines with 
footprints lying between $r_{\rm ovc} = 0.95$ and $r_{\rm ovc} = 1.0$, where 
$r_{\rm ovc}$ are open volume radius coordinates on the polar cap \citep[see e.g. ][]{dyk04}, where 
$r_{\rm ovc} = 0.0$ is the magnetic axis and $r_{\rm ovc} = 1.0$ is the outer rim of the 
polar cap.  The ingoing emission was traced from an outer radius of $r_{\rm max} = 1.2$,
in units of light cylinder radius, $R_{\rm LC} = c/\Omega$, to the neutron star surface, 
in both vacuum retarded dipole \citep{DH04} and force-free \citep{CK10,Har11} magnetic fields.
We find that the resulting emission pattern in observer angle vs. phase with respect to the
rotation axis does indeed show caustics, but the peaks in the light curves do not 
have phase offsets with the outgoing emission peaks that are near 0.25.

\begin{figure*}
\centering
\includegraphics[angle=0,width=10cm]{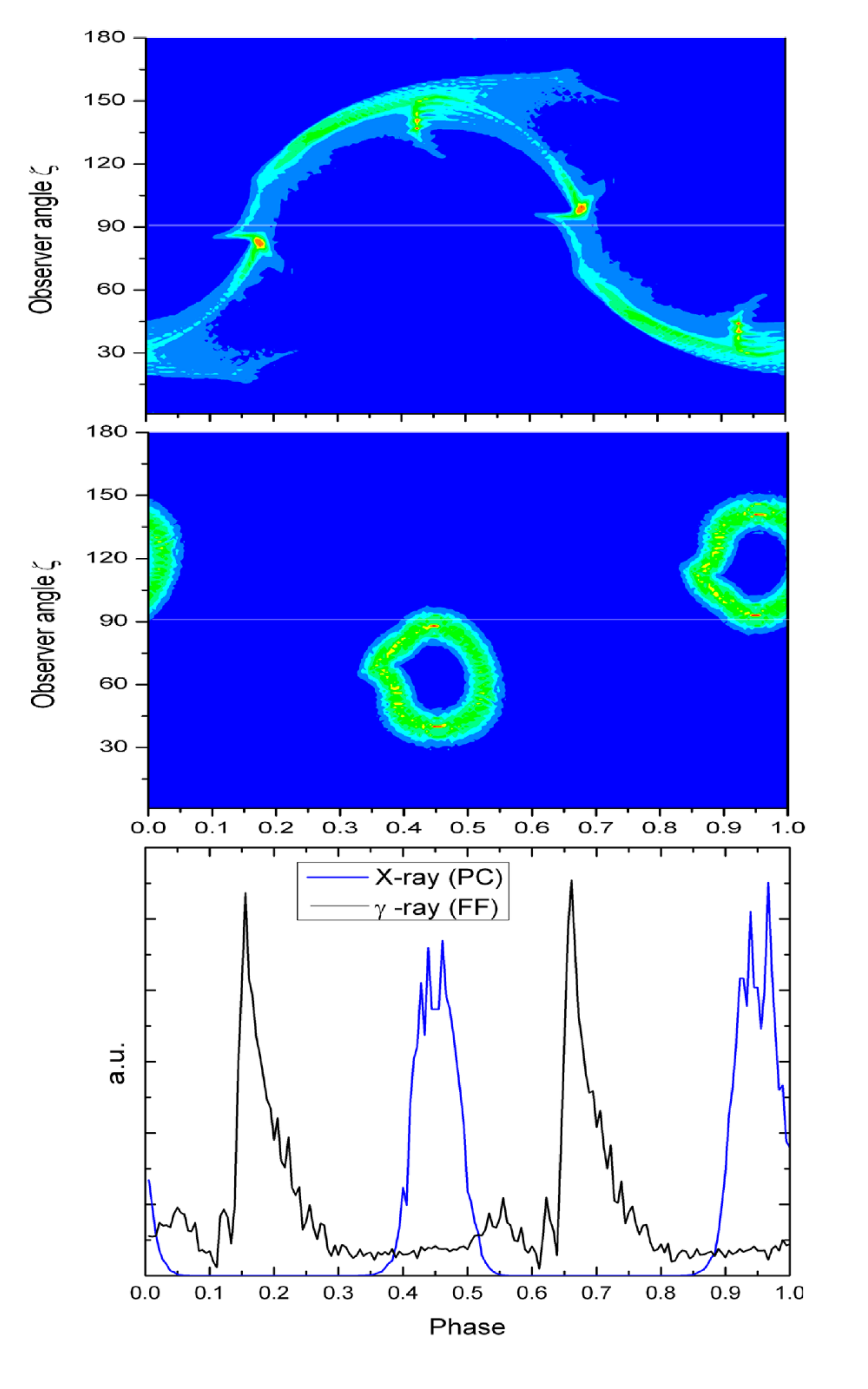}
\protect\caption{{\footnotesize Sky maps of emission in observer angle $\zeta$ vs. phase $\phi$ with respect to the
rotation axis for a magnetic inclination angle $\alpha = 60^\circ$ for a) simulated $\gamma$-ray caustic emission
from the outer magnetosphere for a separatrix layer model in a force-free magnetic field, and b) simulated cone beam
X-ray emission from the polar caps for an emission altitude $r = 0.2 R_{rm LC}$.  c) Model $\gamma$-ray (black) and X-ray
(blue) light curves for a viewing angle of $\zeta = 90^\circ$ (white lines in the skymaps).}
\label{fig-model}}
\end{figure*}

We then explored the possibility that the $\gamma$-ray emission comes 
from the outer magnetosphere and the X-rays comes from the pair cascades above
the polar caps.  We simulated the $\gamma$-ray emission in a force-free magnetosphere,
using a geometry for outward-going emission similar to that of the separatrix model \citep{BS10}, 
between $r_{\rm ovc} = 0.95$ and $r_{\rm ovc} = 0.99$.  The maximum emission radius
was assumed to be $r_{\rm max} = 1.5$, with maximum cylindrical radius of $r_{\rm h} = 2.0$,
so that some emission comes from outside the light cylinder, near the current sheet.  
The photon emission directions are determined entirely in the non-rotating inertial frame, as in \citet{BS10}, 
where the photons are emitted parallel to the particle velocity which is a sum of the drift velocity
and a component parallel to the local magnetic field line.  A sky map of the emission in 
observer angle $\zeta$ vs. phase $\phi$ with respect to the rotation axis for a magnetic 
inclination angle $\alpha = 60^\circ$ is shown in Figure~\ref{fig-model} (cf. \citet{BS10} Figure 9).   
In this map, the magnetic poles are located at $\phi$ = 0, $\zeta$ = 120$^{\circ}$, $\phi$ = 0.5, $\zeta$ = 
60$^{\circ}$.  An observer at $\zeta$ = 90$^{\circ}$ (white horizontal line) will cut through the caustic 
pattern twice to see two peaks at $\phi \sim 0.16$ and $\phi \sim 0.67$, as shown in Figure~\ref{fig-model}.

We simulated the X-ray emission as a cone beam with peak emission just inside 
the polar cap rim $r_{\rm ovc} = 1.0$.  Since the cone beam function modulates the emission 
along all field lines, we allow emission between $r_{\rm ovc} = 0.1$ and $r_{\rm ovc} = 1.2$.  
The cone beam geometry is the same as described in \citet{Stor07}, 
but we allow the altitude of the emission to be a free parameter at a given radius, 
that was adjusted to supply a negative phase shift.  This phase shift, when added to the positive 
phase shift of the $\gamma$-ray peaks can supply the total phase offset between $\gamma$-ray and X-ray peaks.
We find that an X-ray emission altitude of around $r = 0.2 R_{\rm LC}$ is needed to 
give a total phase shift near the observed value.  A sky map of the simulated cone emission
for this case is shown in Figure~\ref{fig-model} for inclination angle $\alpha = 60^\circ$.  An observer at 
$\zeta = 90^\circ$ (white horizontal line) will cut through cones from both magnetic poles
to see two peaks in the light curve.  As shown in Figure~\ref{fig-model}, the phase offset between the X-ray 
and $\gamma$-rays peaks is around 0.24, so this scenario seems promising.  However, the radio cone beam 
emission would occur at lower altitude, forming a smaller cone in the sky map to avoid detection. 
If retarded vacuum dipole field geometry is used instead to simulate the $\gamma$-ray light curve, the first $\gamma$-ray 
peak would lie at phase $\sim 0.08$, and the X-ray cone beam would need to be at an  
implausibly high altitude to make up a total phase offset near 0.25.  The force-free field geometry 
is therefore strongly favored in this scenario.

We note that such a model is in agreement with the empirical results from \citet{mar11}:
as reported, an important difference both in position and height of the X-ray and $\gamma$-ray emitting regions
would fully explain the large spread of the distance-independent X-ray to $\gamma$-ray flux ratio they found.
Moreover, as reported in \citet{mar14}, the alignment between thermal and non-thermal X-ray peaks noted for many pulsars
\citep[see e.g. Geminga, PSR J0659+1414 , PSR J1057$-$5226 and PSR J1741$-$2054;][]{del05,mar14}
further suggests that the non-thermal emission is generated in a region near
the pulsar poles (e.g. in our polar cap emission model). Also, the low X-ray
luminosity of radio-quiet pulsars in the X-ray band \citep{mar11} suggests
that the radio and X-ray emission regions may be in close proximity.
A future, deeper exploration of these pulsar emission characteristics, together with the modelling
of each pulsar, will be able to confirm or rule out our X-ray polar-cap emission model.

The very hard X-ray spectrum (as a comparison, the Crab has a photon
index of 1.6 and Vela of 2.7) and relatively high X-ray flux of J1813
are also unusual for its age. In the $\gamma$-ray-emitting pulsar zoo
only PSR J1811$-$1926, and possibly J2229+6114, have a similar hard
spectrum \citep{mar11}. We note that the three pulsars are quite similar in period, age and energetics. 
Moreover, we did not detect any thermal emission from this pulsar. Very
young pulsars, like the Crab and B1509$-$58, have high levels of
non-thermal X-ray flux relative to $\gamma$-ray flux, and no detected
thermal emission. Middle-aged pulsars, such as Vela and B0656+14, have
dominant thermal emission that is best fit with hot and cool
components, plus a smaller power law component. Given that one expects 
the presence of both heating and cooling thermal emission in a pulsar
of this age (43 kyr), the relative level of the non-thermal emission
is much higher than seen for other middle-aged pulsars. An estimate of
the expected thermal component in J1813 from polar cap heating
from Harding \& Muslimov (2001) is $L_{PC} = 3 \times 10^{31}$ erg s$^{-1}$
from a surface with a $\sim$400 m radius \citep{del05}. From
\citet{pon09} we can also expect a thermal cooling component from the
entire surface $L_{cool} \gtrsim 7 \times 10^{31}$ erg s$^{-1}$. Under the
hypothesis of a pulsar distance of 2.5 kpc, neither component would
be detected due to the high absorption. We can set upper 
limits for the polar cap heating luminosity at $L_{PC} \lesssim 5.6
\times 10^{31}$ erg s$^{-1}$ and thermal cooling luminosity at
$L_{cool} \lesssim 1.3 \times 10^{33}$ erg s$^{-1}$. We conclude that
the lack of detection of thermal emission from J1813 is due to the high absorption.

The non-thermal X-ray flux of J1813 is relatively high, compared to its $\gamma$-ray flux.  In fact, its $\gamma$-to-X flux ratio is 234$\pm$6,
three times less than the lowest one of the radio-quiet pulsar family \citep{mar11,abd13}. We question whether J1813
is a radio-quiet pulsar, or a radio-loud one with its radio counterpart unobservable due to the large distance.
Its upper limit radio flux at 1400 MHz is 17$\mu$Jy \citep{ray11}. J1813 falls just below the
value of 30$\mu$Jy conventionally used to divide radio-quiet and radio-loud pulsars \citep{abd13}, but over the threshold in
pseudo-luminosity of 100 $\mu$Jy - kpc$^2$ for distances $\gtrsim$2.5 kpc. For comparison, J0106+4855 and J1907+0602
are at a distance of $\sim$ 3 kpc and have a radio flux of 8 and 3.4 $\mu$Jy respectively, with $\gamma$-to-X
flux ratios compatible with those of radio-loud pulsar family. On an observational basis, we cannot therefore conclude
that J1813 does not emit in the radio band. We nevertheless stress that emission geometry estimates
by \citet{pie12} point to a lack of radio emission along our line of sight.

From J1813 X-ray (0.3-10 keV) off-pulse spectroscopy we can derive a 3$\sigma$ flux upper limit for a possible nebula of
1.5 $\times$ 10$^{-13}$ erg cm$^{-2}$ s$^{-1}$, with a photon index of 1.25$\pm$0.21 (1$\sigma$ error). A photon index higher
than that of the pulsar is in agreement with theoretical expectations for synchrotron-emitting nebulae.
Such a value corresponds to a 100\% pulsed fraction from the pulsar, so that all the off-pulse emission comes from the nebula.
\citet{kar08} correlate the nebular and non-thermal pulsar X-ray luminosities for all the nebulae detected by {\it Chandra}. 
Our upper limit nebular flux is barely in agreement with the lower bound of their relation.
This would require that most of the unpulsed component of our source comes from the nebula, making the pulsar about 100\% pulsed.
From the analysis of the {\it Chandra} PSF, the nebula must be within
a 1.5$''$ radius of the pulsar. Assuming standard relations \citep{gae06}, the distance between the pulsar and the head of the termination
shock is expected to be $r_s = (\dot{E}/4\pi c\rho_{ISM}v_{psr}^2)^{1/2}$, where $\rho_{ISM}$
is the ambient density and $v_{psr}$ is the pulsar space velocity. For a typical pulsar velocity (500 km s$^{-1}$)
and ambient density (0.1 atoms cm$^{-3}$) at 2.5 kpc, this would place the shock at 5$''$ from the pulsar
(for instance Vela nebula would have a 5$''$ radius at 2.5 kpc). In
any case, we see no sign of a nebula down to 1.5$''$:
this requires a very high pulsar velocity ($>$ 1800 km s$^{-1}$) and/or interstellar medium density ($>$ 1.3 cm$^{-3}$)
and/or pulsar distance ($>$ 9 kpc).

From the J1813 SED we can argue that there is a discontinuity (smooth or sudden) in the photon index
in the hard X-ray band.
The youngest pulsars (Crab, PSR B1509$-$58) usually have a similar SEDs, peaked in hard X-rays
and with a smooth connection between $\gamma$-rays and X-rays \citep{kas06}.
We suggest that J1813 could be younger than its characteristic age ($\tau_c$ = 43 kyr).
Nevertheless, we note that, while the X-ray thermal emission from a 5-10 kyr-old Supernova Remnant (SNR)
would not be detected due to the high absorption column,
we would expect a brighter-than-usual pulsar wind nebula from the interaction of the pulsar
and the SNR \citep[see e.g.][]{buc11}.
Middle-aged pulsars instead have comparatively weak
non-thermal emission in the X-ray band since their power peaks at GeV energies,
and there is a gap in the detected spectrum between the X-ray and $\gamma$-ray bands.
In several cases (Vela, PSR B1055$-$52) an extrapolation between the
two is plausible, but in others \citep[Geminga, PSR B1706$-$44; ][]{got02} a
connection is not clear. In order to discriminate different multi-wavelength emission models,
the spectrum and timing of J1813 in the hard X-ray band would be of the outmost importance.

\section{Conclusions}

We have analyzed our recent, deep {\it XMM-Newton} and {\it Chandra} observations of the energetic radio-quiet PSR J1813$-$1246. We have also
extended the $\gamma$-ray ephemeris to a 5-year period. J1813 had two glitches during this time
period. Its $\gamma$-ray light curve is characterized by two peaks, separated by 0.5 in phase, with a bridge in between.
No off-pulse emission has been detected.
A phase-resolved spectral analysis revealed a change in the photon index, with a softening during the peaks.
The X-ray spectrum is non-thermal, harder than all the other {\it Fermi} pulsars ($\Gamma$=0.85$\pm$0.03)
and highly absorbed (N$_H$=1.56$\pm$0.07 $\times10^{22}$ cm$^{-2}$).
Detection of thermal emission (from hot spots and from cooling) was not expected due to the high absorption column.
Based on such absorption, on the analysis of serendipitous sources around the pulsar and on
radio observations of the numerous dark clouds in the J1813 region, we 
propose that J1813 is more than 2.5 kpc distant.
Such a large distance would make faint radio pulsations undetectable, even if geometrical models point towards a radio-quiet pulsar.
We also found a hint of detection in the {\it INTEGRAL} IBIS/ISGRI band (30-500 keV), that perfectly matches the Spectral
Energy Distribution of J1813. We detected X-ray pulsations with very high confidence, with a light curve
characterized by two sharp, asymmetrical peaks,
separated by 0.5 in phase. The X-ray peaks lag the $\gamma$-ray ones by 0.25 in phase.
The pulsed fraction of the X-ray source is 96$\pm$3\%, with a faint off-pulse emission detected
that can be due to nebular emission, that is nevertheless undetectable through brightness profile analysis down to
1-1.5$''$. A phase-resolved spectral analysis revealed no significant X-ray spectral changes during the pulsar phase.

Outer gap and slot gap models predict shapes similar to our $\gamma$-ray profile for
very high observer angles to the spin axis (87$^{\circ}$ and 78$^{\circ}$ respectively). 
In high-altitude emission models particle acceleration and emission along the last open trailing field lines up
to near the light cylinder is expected to produce non-thermal $\gamma$-ray, X-ray and optical pulses at the same phase in the light
curve. The fact that the observed non-thermal X-ray peaks in J1813
are not only not in phase with the $\gamma$-ray peaks, but are both out of phase by about one
quarter of a period, is not in agreement with both X-ray and $\gamma$-ray emission being outgoing emission from 
the outer magnetosphere.  It is possible that the $\gamma$-ray emission comes from the outer magnetosphere and 
the X-ray emission comes from the polar cap, but at an altitude of about 40 neutron star radii.  The phase offset between 
$\gamma$-ray and X-ray peaks requires the use of force-free magnetic field geometry in modeling the light curves.  X-ray emission from 
polar cap pair cascades is mostly synchrotron radiation from secondary electron-positron pairs that are produced with
a broad spectrum of energies \citep{DH82}.  The emission can extend from a few tenths of a keV up to 10 -100 MeV \citep{DR99,RD99}, 
so could plausibly explain the spectrum of J1813. If this is the 
case, it would be the first time that a clear emission from the polar cap pair cascades has been observed.  

\section{Appendix} \label{appendix}

Based on the study of spectra and possible optical counterparts,
we can classify serendipitous sources as AGNs or candidate stars,
allowing us to constrain the pulsar distance. 
Indeed, after selecting candidate AGNs in the FOV, it is possible to measure from their spectra
the total Galactic column density in the direction of J1813.

As a first step, we studied the brightness profile of each source and compared it with the theoretical (for {\it XMM-Newton} data) or simulated
(for {\it Chandra} data) PSF. All the detected sources are point-like.
We performed a standard {\it XMM-Newton} and {\it Chandra} spectral analysis for the nine sources detected at $>10\sigma$ and with
more than 300 {\it XMM-Newton} net total counts.
The main discriminator among different classes of X-ray-emitting objects is the spectral shape.
The spectra were fitted either with an absorbed power law, well-suited for
AGN, and absorbed double {\tt apec}s (emission spectrum from collisionally-ionized diffuse gas), well-suited for stellar coronae.
From studies on serendipitous X-ray sources in {\it Chandra} and {\it XMM-Newton} observations
\citep[see e.g.][]{nov09,ebi02} the detection probability for other X-ray emitting source classes
in our mid-Galactic-latitude {\it XMM-Newton} observation is negligible.
Four out of the nine considered sources can be fitted only by an absorbed power law (sources \#4,\#5,\#8,\#10),
prompting their AGN classification, and two only by double apecs
(sources \#7,\#11), suggesting a stellar classification, while three remained unclassified.

\begin{figure*}
\centering
\includegraphics[angle=0,width=15cm]{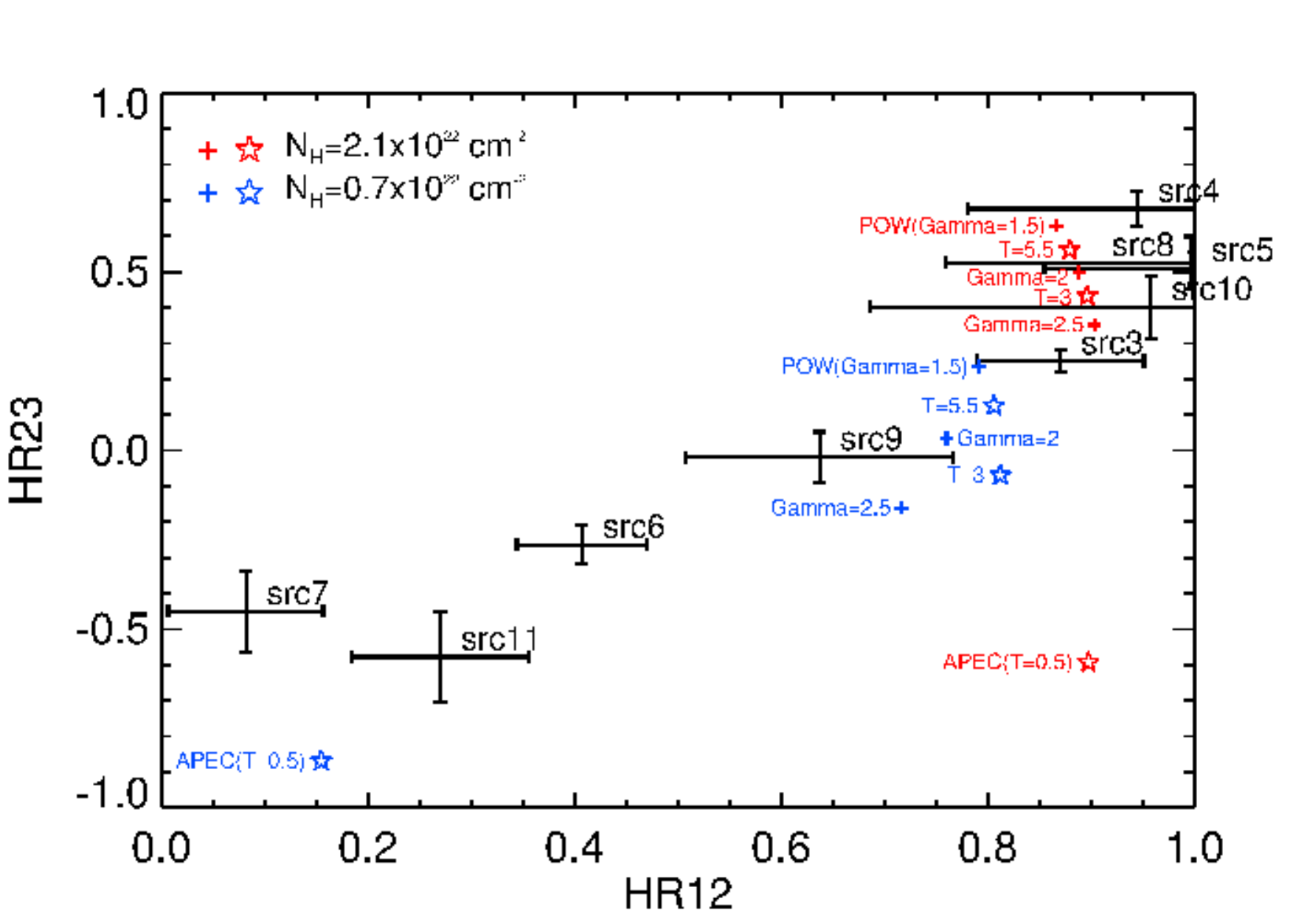}
\protect\caption{{\footnotesize Distribution of HR12 vs. HR23 of the nine selected X-ray sources.
Error bars are reported at 1$\sigma$. Crosses indicate the expected HR12 vs.
HR23 computed for power law spectra with $\Gamma$ from 1.5 and 2.5.
Stars indicate the expected HR12 vs. HR23 computed for apec spectra with kT
from 0.5 to 5.5 keV. Each spectral model is computed using the average interstellar
medium absorption given by \cite{dic90} (red) and thrice that value (blue).}
\label{hr}}
\end{figure*}

Since the count statistics of some of the selected X-ray sources 
is too low to discriminate the spectral model, we
performed a qualitative spectral analysis using the count rate (CR). We measured it
in the three energy ranges (\textit{soft}: 0.3--1 keV; \textit{medium}: 
1--2 keV; \textit{hard}: 2--10 keV) to compute two different Hardness Ratios (HRs):
\begin{dmath}
HR12 = [CR(1-2)-CR(0.3-1)]/[CR(1-2)+CR(0.3-1)]\\
HR23 = [CR(2-10)-CR(1-2)]/[CR(2-10)+CR(1-2)]
\end{dmath}
Adopting the above definition, sources with a small/large HR12 value are little/very absorbed,
while sources with a small/large HR23 value are characterized by a soft/hard spectrum.\\
Figure \ref{hr} shows the distribution of the HRs of the nine serendipitous X-ray sources.
To obtain a further indication on the spectra of the sources, we
compared the measured HRs with the expected ones computed
for two different template spectral models, namely: a power law, 
with photon indexes $\Gamma$ increasing from 1.5
to 2.5, and an apec, with
temperatures kT increasing from 0.5 to 5.5 keV. Each spectral model 
is computed using the average interstellar medium absorption
given by \citet{dic90} and thrice that value (that is the highest value fitted
among the serendipitous sources spectra). The values of the expected HRs 
are overplotted in Figure~\ref{hr}.\\
Sources \#6,\#7 and \#11 are little absorbed and are characterized by a 
rather soft spectrum ($HR12<0.5$ and $HR23<0$),
pointing to a (nearby) star classification. 
Sources \#4, \#5, \#10 and \#8 are probably situated farther than the dark cloud
for their high absorption; their hard spectra suggest that they are likely AGN.
Such a method confirms the spectral results, also adding source 6 to the pool of stars.

Another common way to confirm X-ray classification of sources is based on multi-wavelength
analysis: the X-to-optical flux ratio is a good indicator of the nature of X-ray emitters.
According to \citet{lap06}, AGN have typical logarithms of X-to-optical flux ratios higher than $-0.2$, while stars lower than +1.0.
Inside each X-ray source error box, we looked for
association with optical sources from the NOMAD catalogue \citep{zac05},
considering the V-band magnitude as reference. When the V-band
magnitude was not available for the candidate counterpart, we alternatively used the
R magnitude.
In the case of J1813 field, however, we do not expect to find the optical counterparts of AGN
due to the surprisingly high value of our column densities: in fact we expect magnitudes
well above the NOMAD upper limit of m=21. Few of our AGN-like objects
have optical counterparts inside their error box that could be due to spurious matches.
In order to estimate the number of spurious matches, we used the relation
from \citet{sev05,nov09}. This yielded a probability of chance coincidence of 21\%, wich means that,
at our limiting magnitudes, contamination effects cannot be ignored.
Each of the star-like objects has an optical counterpart
that agrees with the expected X-ray-to-optical flux ratio. Table~\ref{tab2} reports the
associated optical counterparts and expected upper limits for AGN.

Thus, we identified four out of the nine sources we considered as AGN
and three of them as stars. The remaining two objects have possible star-like optical counterparts which yield
reasonable X-ray-to-optical flux ratios. The faintness of the optical counterparts precludes any further analysis.

{\bf Acknowledgments}

We warmly thank Paizis Adamantia, Andrea Giuliani, Fabio Gastaldelli and Andrea Belfiore for all the discussions and helps.
We also thank Massimiliano Razzano and Lemoine-Goumard for their good work as Galactic Coordinators.\\
The $Fermi$ LAT Collaboration acknowledges support from a number of agencies and institutes for both development and the operation of the LAT as well as scientific data analysis. These include NASA and DOE in the United States, CEA/Irfu and IN2P3/CNRS in France, ASI and INFN in Italy, MEXT, KEK, and JAXA in Japan, and the K.~A.~Wallenberg Foundation, the Swedish Research Council and the National Space Board in Sweden. Additional support from INAF in Italy and CNES in France for science analysis during the operations phase is also gratefully acknowledged.
Support for this work was provided by the National Aeronautics and
Space Administration through {\it Chandra} Award Number GO3-14053X
issued by the Chandra X-ray Observatory Center, which is operated by
the Smithsonian Astrophysical Observatory for and on behalf of the 
National Aeronautics Space Administration under contract NAS8-03060.
This work was supported by the ASI-INAF contract I/037/12/0,
art.22 L.240/2010 for the project $''$Calibrazione ed Analisi del satallite NuSTAR$"$.

Facilities: CXO (ACIS), XMM (EPIC), Fermi (LAT), Suzaku (XIS).

\clearpage

\begin{landscape}
\begin{table}
\small
\begin{center}
\caption{J1813 ephemeris\label{tab1}}
\begin{tabular}{ccccccccc}
\tableline\tableline
Validity period & F0 & F1 & F2 & F3 & F4 & F5 & H-value & $\Delta$Phase\tablenotemark{a}\\
MJD & Hz & $10^{-12}$Hz$^2$ & $10^{-22}$Hz$^3$ & $10^{-30}$Hz$^4$ & $10^{-37}$Hz$^5$ & $10^{-44}$Hz$^6$ & - & -\\
\tableline
54682-55114 & 20.80104237955 & -7.93199 & -78.201 & -121.557 & -9.347 & 0 & 720 & -\\
55114-56290 & 20.80107402437 & -7.63091 & -7.966 & -52.707 & -20.321 & -3.85 & 1617 & -0.037$\pm$0.005\\
56290-56508 & 20.80107410361 & -7.62139 & -6.1725 & -1.753 & 0 & 0 & 235 & 0.345$\pm$0.007\\
\tableline
\end{tabular}
\tablenotetext{a}{$\Delta$Phase=Phase$_{fin}$-Phase$_{in}$ is evaluated using the phase lag between the two maxima; a positive value implies a shift of the peaks towards right in the light curve.}
\tablecomments{Ephemeris of J1813 at MJD$_0$=56362.0, corresponding to the X-ray observation epoch. High level derivatives have been added in order to improve the H-value(20 harmonics), so that they are not related to physical values. The obtained H-values are proportional to the number of photons considered ($\propto$ validity period).}
\end{center}
\end{table}
\end{landscape}

\begin{table}
\begin{center}
\caption{Analysis of serendipitous sources\label{tab2}}
\begin{tabular}{cccccc}
\tableline\tableline
source & J2000 coord & N$_H$ & spectrum & HR & $log(\frac{f_{X}}{f_{V}})$\\
- & - &$10^{22}\,cm^{-2}$ & - & - & - \\
\tableline
3& 273.5839 -12.7397\tablenotemark{a} & 0.91$\pm$0.15/0.88$\pm$0.14 & ? & ? & -0.69\\
4& 273.2630 -12.8186\tablenotemark{b} & 2.1$\pm$0.4 & AGN & AGN & $m_R>$25.4\tablenotemark{d}\\
5& 273.2210 -12.6814\tablenotemark{a} & 1.7$\pm$0.3 & AGN & AGN & $m_R>$23.9\tablenotemark{d}\\
6& 273.3608 -12.8407\tablenotemark{b} & 0.40$\pm$0.08/0.98$\pm$0.20 & ? & star & -1.93/-2.32/-0.94\\
7& 273.3553 -12.7529\tablenotemark{b} & 0.81$\pm$0.15 & star & star & -2.50\tablenotemark{c}\\
8& 273.3029 -12.6696\tablenotemark{a} & 1.7$\pm$0.6 & AGN & AGN & $m_R>$24.9\tablenotemark{d}\\
9& 273.5829 -12.7871\tablenotemark{a} & 0.44$\pm$0.18/0.33$\pm$0.16 & ? & ? & -1.62\tablenotemark{c}\\
10& 273.4600 -12.7828\tablenotemark{a} & 1.1$\pm$0.6 & AGN & AGN & $m_R>$22.9\tablenotemark{d}\\
11& 273.4489 -12.7895\tablenotemark{a} & 0.84$\pm$0.13 & star & star & -3.43\tablenotemark{c}\\
\tableline
\end{tabular}
\tablenotetext{a}{Position obtained by {\it XMM-Newton}; typical 90\% error box of 5$''$}
\tablenotetext{b}{Position obtained by {\it Chandra}; typical 90\% error box of 2$''$}
\tablenotetext{c}{For these sources a proper motion has been detected and reported in the NOMAD catalog.}
\tablenotetext{d}{For the high fitted absorption column of AGN-like sources we reported the expected
observable magnitude, based on \citet{lap06}. Such counterparts are not detectable in the NOMAD catalog.}
\tablecomments{Results of the serendipitous sources analysis. Here we report the best X-ray position, the classification
following the spectral and HR methods described in Section~\ref{appendix} and the logarithm of the X-to-optical flux ratio (for all the possible
optical counterparts).
We also report the best fitted column density of the source; if the spectrum is well fitted both by powerlaw and apecs,
we report both the values, respectively.
The X-ray flux is unabsorbed and in the 0.3-10 keV energy range. The optical flux is unabsorbed and
in the V band, from the NOMAD catalog. The errors on column densities are at a 90\% confidence.}
\end{center}
\end{table}

\clearpage


\begin{thebibliography}{}
\bibitem[Abdo et al.(2009a)]{abd09} Abdo, A.A., Ackermann, M, Ajello, M. et al. 2009, Science, 325, 840 
\bibitem[Abdo et al.(2009b)]{abd09b} Abdo, A.A., Ackermann, M, Ajello, M. et al. 2009, ApJ, 183, 46 
\bibitem[Abdo et al.(2013)]{abd13} Abdo, A.A., Ajello, M., Allafort, A. et al. 2013, ApJS, 208, 17 
\bibitem[Abergel et al.(2014)]{abe14} Abergel, A., Ade, P.A.R., Aghanim, N., Alina, D., Alves, M.I.R. et al., 2014, arXiv 1312.1300
\bibitem[Bai \& Spitkovsky(2010)]{BS10} Bai, X.-N., \& Spitkovsky, A. 2010, ApJ, 715, 1282
\bibitem[Bevington(1969)]{bev69} Bevington P.R. 1969, Data Reduction and Error Analysis for the Physical Sciences (New York: McGraw-Hill)
\bibitem[Bucciantini et al.(2011)]{buc11} Bucciantini, N., Arons, J, Amato E., 2011, MNRAS, 410, 381
\bibitem[Caraveo et al.(2004)]{car04} Caraveo, P.A., De Luca, A., Mereghetti, S., Pellizzoni, A., Bignami, G. F., 2004, Sci, 305, 376
\bibitem[Caraveo(2013)]{car13} Caraveo, P.A., 2013, ArXiv:1312.2913	
\bibitem[Cash(1979)]{cas79} Cash, W., 1979, ApJ, 228, 939
\bibitem[Contopoulos \& Kalapotharakos(2010)]{CK10} Contopoulos, I. \& Kalapotharakos, C. 2010, MNRAS, 404, 767
\bibitem[Cheng et al.(1986)]{che86} Cheng, K.S., Ho, C., Ruderman, M. 1986, ApJ, 300, 500
\bibitem[Dame et al.(2001)]{dam01} Dame, T.M., Hartmann, D., Thaddeus, P., 2001, ApJ, 547, 792
\bibitem[Daugherty \& Harding(1982)]{DH82}Daugherty, J. K. \& Harding, A. K. 1982, ApJ, 252, 337
\bibitem[De Jager \& Busching(2010)]{dej10} De Jager, O. \& Busching, I., 2010, A\&A, 517, 9
\bibitem[De Luca et al.(2005)]{del05} De Luca, A., Caraveo, P.A., Mereghetti, S., Negroni, M., Bignami, G.F. 2005, ApJ, 623, 1051 
\bibitem[Dobashi(2011)]{dob11} Dobashi, K. 2011, PASJ, 63, 1
\bibitem[Dickey \& Lockman(1990)]{dic90} Dickey, J.M. \& Lockman, F.J., 1990, ARA\&A, 28, 215
\bibitem[Dyks \& Rudak(2003)]{dyk03} Dyks, J. \& Rudak, B. 2003, ApJ, 598, 1201
\bibitem[Dyks \& Rudak(1999)]{DR99} Dyks, J. \& Rudak, B., 1999, ApL\&C, 38, 41
\bibitem[Dyks et al.(2004)]{dyk04} Dyks, J., Harding, A.K.. Rudak, B., 2004, ApJ, 606, 1125
\bibitem[Dyks, J. \& Harding, A. K.(2004)]{DH04} Dyks, J. \& Harding, A.K. 2004, ApJ, 614, 869
\bibitem[Ebisawa et al.(2002)]{ebi02} Ebisawa, K., Paizis, A., Maeda, Y. et al. 2002, bhap.conf, 25
\bibitem[Gaensler \& Slane(2006)]{gae06} Gaensler, B.M. \& Slane, P.O., 2006, ARA\&A, 44, 17
\bibitem[Garmire et al.(2003)]{gar03} Garmire, G.G., Bautz, M.W., Ford, P.G. et al. 2003, SPIE 4851, 28
\bibitem[Gotthelf et al.(2002)]{got02} Gotthelf, E.V., Halpern, J.P., Dodson, R., 2002, ApJ, 567, 125
\bibitem[Harding \& Muslimov(2004)]{har04} Harding, A.K. \& Muslimov, A.G., 2004, 35th COSPAR Scientific Assembly, 35, 562
\bibitem[Harding et al.(2008)]{har08} Harding,A.K., Stern, J.V., Dyks, J., Frackowiak, M., 2008, ApJ, 680, 1378
\bibitem[Harding(2013)]{har13} Harding, A.K., 2013, JASS, 30, 145
\bibitem[Harding et al.(2011)]{Har11} Harding, A. K., DeCesar, M. E., Miller, M. C., Kalapotharakos, C., \& Contopoulos, I. 2011, arXiv:1111.0828
\bibitem[Hou et al.(2009)]{hou09} Hou, L.G., Han, J.L., Shi, W.B., 2009, A\&A, 499, 473
\bibitem[Kalapotharakos et al.(2012)]{kal12} Kalapotharakos, C., Harding, A.K., Kazanas, D., Contopoulos, I., 2012, ApJ, 754, 1
\bibitem[Kalberla et al.(2005)]{kal05} Kalberla, P.M. Burton, W.B., Hartmann, D. 2005, A\&A, 440, 775
\bibitem[Kargaltsev \& Pavlov(2008)]{kar08} Kargaltsev, O. \& Pavlov, G. G., 2008, AIPC, 983, 171
\bibitem[Kaspi et al.(2006)]{kas06} Kaspi, V.M., Roberts, M.S.E., Harding, A.K., 2006, Compact stellar X-ray sources, edited by Walter Lewin \& Michiel van der Klis, Cambridge Astrophysics Series, No. 39, 279
\bibitem[Kerr(2011)]{ker11} Kerr, M. 2011, ApJ, 732, 38
\bibitem[Kunz\&Snowden(2008)]{kun08} Kuntz, K.D. \& Snowden, S.L. 2008, A\&A, 478, 575
\bibitem[Kuster(1999)]{kus99} Kuster, M., Benlloch, S., Kendziorra, E., Briel, U.G., 199, Proc. SPIE Vol. 3765, p. 673-682
\bibitem[La Palombara et al.(2006)]{lap06} La Palombara, N., Mignani, R.P., Hatziminaoglou, E. et al. 2006, A\&A, 458, 245
\bibitem[Lebrun et al.(2003)]{leb03} Lebrun, F., Leray, J.P., Lavocat, P., et al. 2003, A\&A, 411, 141
\bibitem[Li et al.(2012)]{li12} Li, J., Spitkovsky, A., Tchekhovskoy, A., 2012, ApJ, 746, 60
\bibitem[Marelli et al.(2011)]{mar11} Marelli, M., De Luca, A., Caraveo, P.A. 2011, ApJ, 733, 82
\bibitem[Marelli et al.(2013)]{mar13} Marelli, M., De Luca, A., Salvetti, D. et al. 2013, ApJ, 765, 36
\bibitem[Marelli et al.(2014)]{mar14} Marelli, M., Belfiore, A., Saz Parkinson, P. et al. 2014, arXiv:1404.1532
\bibitem[Mattox at al.(1996)]{mat96} Mattox, J.R., Bertsch, D.L., Chiang, J. et al. 1996, ApJ, 461, 396
\bibitem[McClure et al.(2009)]{mcc09} McClure-Griffiths, N.M., Pisano, D.J., Calabretta, M.R., Ford, H.A., Lockman, F.J. et al. 2009, ApJS, 181, 398
\bibitem[Mitsuda et al.(2007)]{mit07} Mitsuda, K., Bautz, M., Inoue, H. 2007, PASJ, 59, 1
\bibitem[Novara et al.(2009)]{nov09} Novara, G., La Palombara, N., Mignani, R.P. et al. 2009, A\&A, 501, 103
\bibitem[Nolan et al.(2012)]{nol12} Nolan, P.L., Abdo, A. A., Ackermann, M. et al. 2012, ApJ, 199, 31
\bibitem[Pierbattista et al.(2012)]{pie12} Pierbattista, M. Grenier, I.A., Harding, A.K., Gonthier, P.L. 2012, A\&A, 545, 42
\bibitem[Pierbattista et al.(2014)]{pie14} Pierbattista, M., Harding, A.K., Grenier, I.A. 2014, arXiv:1403.3849, submitted to A\&A
\bibitem[Pons et al.(2009)]{pon09} Pons, J.A., Miralles, J.A., Geppert, U. 2009, A\&A, 496, 207
\bibitem[Protassov et al.(2002)]{pro02} Protassov, R., Van Dyk, D.A., Connors, A., Kashyap, V.L., Siemiginowska, A. 2002, ApJ, 571, 545
\bibitem[Ray et al.(2011)]{ray11} Ray, P.S., Kerr, M., Parent, D. et al. 2011, ApJ, 194, 17
\bibitem[Read(2004)]{rea04} Read, A.M. 2004, XMM-CCF-REL-167
\bibitem[Romani \& Yadigaroglu(1995)]{rom95} Romani, R.W. \& Yadigaroglu, I.A., 1995, ApJ, 438, 314
\bibitem[Rudak \& Dyks(1999)]{RD99} Rudak, B. \& Dyks, J., 1999, MNRAS, 303, 477
\bibitem[Saz Parkinson et al.(2010)]{saz10} Saz Parkinson, P. M., Dormody, M., Ziegler, M. et al. 2010, ApJ, 725, 571
\bibitem[Severgnini et al.(2005)]{sev05} Severgnini, P., Della Ceca, R., Braito, V. et al. 2005,  A\&A, 431, 87
\bibitem[Skrutskie et al.(2006)]{skr06} Skrutskie, M.F., Cutri, R.M., Stiening, R. et al. 2006, AJ, 131, 1163
\bibitem[Spitkovsky(2006)]{spi06} Spitkovsky, A., 2006, ApJ, 648, 51
\bibitem[Story et al.(2007)]{Stor07} Story, S. A., Gonthier, P. L. \& Harding, A. K. 2007, ApJ, 671, 713
\bibitem[Struder et al.(2001)]{str01} Struder, L., Briel, U., Dennerl, K. et al. 2001, A\&A, 365, L18
\bibitem[Takata \& Chang(2007)]{tak07} Takata, J. \& Chang H.K., 2007, ApJ, 670, 677
\bibitem[Timokhin(2006)]{tim06} Timokhin, A.N., 2006, MNRAS, 368, 1055	
\bibitem[Timokhin \& Harding(2014)]{TH14} Timokhin, A. \& Harding, A. K. 2014, in preparation.
\bibitem[Treves et al.(2001)]{tre01} Treves, A., Popov, S.B., Colpi, M., Prokhorov, M.E., Turolla, R., 2001, ASPC, 234, 225
\bibitem[Turner et al.(2001)]{tur01} Turner, M.J.L., Abbey, A., Arnaud, M. et al. 2001, A\&A, 365, 27
\bibitem[Wang et al.(2013)]{wan13} Wang, Z., Breton, R.P., Heinke, C.O., Deloye, C.J., Zhong, J. 2013, ApJ, 765, 151
\bibitem[Watters \& Romani(2011)]{wat11} Watters, K.P. \& Romani, R.W. 2011, ApJ, 727, 123
\bibitem[Zacharias et al.(2005)]{zac05} Zacharias, N., Monet, D.G., Levine, S.E. et al. 2005, yCat, 1297, 0
\end{thebibliography}
\end{document}